\documentclass[12pt,a4paper]{article}
\usepackage{jheppub_kim}
 \usepackage{slashed}
\usepackage{longtable}
\usepackage{epsfig}
\usepackage{amssymb,amsmath}
\usepackage{amsthm}
\usepackage{graphicx}
 \usepackage{graphics}
\usepackage[hang,nooneline,scriptsize]{subfigure}
\usepackage{subfigure}
\usepackage{array}
\usepackage{xcolor,soul}
\usepackage{wasysym}
\begin{document}
  
\title{\textcolor{blue}{Weak Deflection Angle, Hawking Radiation, Greybody Bound and Shadow Cast for Static Black Hole in the Framework of $f(R)$ Gravity}}
 
\author[a,b] {Surajit Mandal\footnote{Corresponding author}} 

\affiliation[a]{Department of Physics, Jadavpur University, Kolkata, West Bengal 700032, India. }
\affiliation[b]{Department of Physics, AKPC Mahavidyalaya, Bengai, West Bengal 712611, India.}

\emailAdd{surajitmandalju@gmail.com; smobdal098@gmail.com}

\abstract{In this work, we probe the weak gravitational lensing by a static spherically symmetric black hole in view of $f(R)$ gravity in the background of the non-plasma medium (vacuum). We provide a discussion on a light ray in a static black hole solution in $f(R)$ gravity. To adore this purpose, we find the Gaussian optical curvature in weak gravitational lensing by utilizing the optical geometry of this black hole solution. Furthermore, we find the deflection angle up to the leading order by employing the Gauss-Bonnet theorem. We present the graphical analysis of the deflection angle with respect to the various parameters that govern the black hole. Further, we calculate the Hawking temperature for this black hole via a topological method and compare it with a standard method of deriving the Hawking temperature. We also analyze the Schr$\ddot{o}$dinger-like Regge-Wheeler equation and derive bound on the greybody factor for a static black hole in the framework of $f(R)$ gravity and graphically inquire that bound converges to 1. We also investigate the silhouette (shadow) generated by this static $f(R)$ black hole. Moreover, we constrain the non-negative real constant and cosmological constant from the observed angular diameters of M$87^{\star}$ and Sgr $A^{\star}$ released by the EHT. We then probe how cosmological constant, non-negative real constant and mass affected the radius of shadow. Finally, we demonstrate that, in the eikonal limit, the real part of scalar field quasinormal mode frequency can be determined from the shadow radius.
}	
	\maketitle 

\section{Introduction}
In the theory of general relativity (henceforth GR), black holes are one of the most attractive and engrossing objects in our universe. Black holes provide a cabalistic tool to test and verify the essential laws of the universe. Initially, the image of a black hole was only in simulations, but later the Event Horizon Telescope (EHT) collaboration \cite{m1,m2,m3,ligo4} exposed the first picture of the black hole giving confirmation of the existence of it. In the universe, a black hole is a massive object having exorbitant gravity that nothing (including light) can escape from its pull. The appearance of gravitational lensing and gravitational waves as a consequence of GR discovered by Einstein in 1916 \cite{m4}. In addition, the gravitational waves from black holes and neutron stars merger have been determined via the Laser Interferometer Gravitational-Wave Observatory (LIGO), which corroborated the theoretical prospect that nicely impels with experimental studies \cite{ligo1,ligo2}. A spacious variety of gravity theories encounter many flaws after the identification of gravitational waves, meanwhile, the identification of gravitational waves also seized the deliberation in the field of gravitational lensing (abbreviated as GL) \cite{i5}. In 1801, Soldner first introduced the concept of GL within the context of Newton’s theory of gravitation \cite{i6}. GL is an enormously useful way of appreciating the knowledge of galaxies and the universe. During the period of solar eclipse, the bending within the starlight appears as a consequence of the Sun’s gravitational field. After this, they noticed that they got findings similar to the results predicted by Einstein. As a result, it played a documentary role in the experiential warrant of Einstein’s prediction.  Gravitational lensing denotes the bending of light around extensive objects which is predicted by GR. Three types of GL can be found in literature such as strong GL, weak GL, and micro GL which have been documented in Refs. \cite{i7,i8,i9,i10,i11,i12}.

Weak lensing is a feasible way for the identification of massive astronomical object (including black hole) which has huge radii and mass without knowing about their structures. Gravitation lensing also studies the expansion of the universe. Nowadays, the study of GL has become an interesting area for researchers. GL plays an important role in investigating many astrophysical objects like black holes, naked singularities, and wormholes (see e.g. Refs. \cite{i7,i8,i9,i10,i11,i12,i13,i14,i15,i16,i17,i18}). Many investigations, after the confirmation of Gibbons and Werner's about the feasible manner to derive the deflection angle from black holes, have connected GL with the Gauss-Bonnet theorem (henceforth GBT). The deflection angle is calculated by integrating the Gaussian optical curvature of the black hole \cite{i19}, which is articulated as follows:
\begin{equation}\label{78}
\boldsymbol{\tilde\delta}=- \iint_{\mathcal{D}_{\infty}} \mathcal{K} d S
\end{equation}
here $\tilde\delta$, $\mathcal{K}$, and  $dS$ are the deflection angle, the Gaussian optical curvature, and the optical surface respectively. Also, $\mathcal{D}_{\infty}$ denotes the infinite domain encompassed by the photon ray. Thus, the Gibbons and Werner methodology has addressed in a unique prospect for black holes and wormholes \cite{i20,i22,i23,i24,i24a,i25,i26,i27,i28,i29,i30,i31,i32,i33,i34,i35}.

The shadow is essentially illustrated as the two-dimensional ($2D$) dark region occurs in the celestial plane. The concept of the black-hole
shadow appears when there exists a geometrically thick and optically thin emission region around the event horizon of a black hole \cite{sha}. The size of the shadow radius depends on its contour and on the intrinsic parameters of the black hole. Furthermore, the shadow size is estimated by the instability (orbital) of the photon rays from the photon sphere, which merely seems to be a dark, $2D$ region for a faraway observer exposed by its uniform, bright surrounding \cite{m1,i36,i37,i38,i39,i41,i42,i43,i44}. Apart from these investigations, several significant studies on constraining the fundamental parameters of the theories against observational data from M$87^{\star}$ and Sgr $A^{\star}$ have been conducted in Refs.\cite{sunny1,sunny2,sunny3,sunny5,sunny6,konoplya}.

Generally, when a black hole is perturbed, it undergoes a damping oscillation which is further decorated by a superposition of exponentially decaying sinusoidal modes known as quasinormal modes (QNMs) \cite{f20,qnm1}. During the oscillations, the black hole starts expending its energy by emitting gravitational waves forcing the whole system to become a dissipative system and this is why the notion \textit{quasi} appears. The QNMs frequencies are in general complex-valued functions in which the real part corresponds to the oscillation and the imaginary part describes its decay. Moreover, the imaginary part of the QNMs gives information about the stability of the black hole. It is well-established, in GR, that the eikonal QNMs for most of the static, spherically symmetric, and asymptotically flat black holes can be directly obtained via the shadow radius of the black holes \cite{qnm2}. It would be interesting to study the correspondence between the shadow radius and the real part of QNMs in eikonal limit for static $f(R)$ black hole.

In the framework of quantum field theory, Hawking's discovery tells us that particles can escape from black holes known as radiation \cite{i45}. This becomes one of the exigent findings of Hawking and later this radiation is familiar as Hawking radiation. Two distinct ideas, such as GR and quantum mechanics were united by his work \cite{i45,i46}. In view of quantum field theory, there is a possibility to create and annihilate the particles. When pair production occurs in the vicinity of a black hole’s horizon, it is noteworthy that one of the particles from that pair production escapes from the black hole resulting in Hawking radiation, while the rest of the particles fall back into the black hole. In GR, spacetime may bend around massive objects (like a black hole), producing a gravitational potential inside which the particles can move. The black hole can transmit some of the radiation and goes to infinity and the remaining are reflected back into the black hole. As a result, the observed Hawking radiation which passes through the gravitational potential and which has not been crossed through the gravitational potential is different. The difference has been investigated by the “greybody factor”.

Several methods have been established for calculating Hawking radiation \cite{gws,gws1,gws2,gws3,gws4}. Hawking and Gibbons \cite{gws} have derived Hawking temperature by utilizing the Euclidean path integral for the gravitational field. Later, Robson, Villari, and Biancalana (RVB) \cite{f13,f14} proposed that Hawking temperature of a black hole can also be attained topologically. This topological technique requires
the invariants of the topology like Euler characteristic\footnote{The Euler characteristic of Euclidean geometry is represented by the parameter $\chi$ and it can give the number of the Killing horizons. Moreover, it is related to the structure of the manifold and it is a topological invariant \cite{f13,f14}.} and GBT. One can interpret the Hawking radiation in a crystallized way for the 2-dimensional spacetime of Euclidean geometry without losing the information of 4-dimensional spacetime under the consideration of topological technique. In the framework of the topological method, Zhang et. al \cite{hww} studied the Hawking temperature of the BTZ black hole. Moreover, few studies to derive the Hawking temperature for black holes by employing the topological strategy can be found in Refs. \cite{hww1,f17}. In the context of non-linear electrodynamics, the exploration of the Hawking temperature for a magnetically charged black hole through surface gravity and horizon was performed in Ref. \cite{hww2}.

The greybody factor can be estimated by using several methods such as the matching technique \cite{i47,i48,i49}, and  WKB approximation \cite{i50,i51}. People developed a new technique to compute the greybody factor with accuracy. This new method includes the computation of the rigorous bound on the greybody factor. Many researchers, with the help of this new technique, have estimated the bounds on the greybody factor for the scenario of different black holes \cite{i52,i53,f12a}. The greybody factor of a charged massive BTZ black hole is discussed in Ref. \cite{f12a}. Apart from these studies, there are several significant studies dealing with rigorous bound on the greybody factor for different black hole systems \cite{i47,gdy2,gdy3,gdy4,gdy5,gdy6,gdy7,gdy8,gdy9}. Motivated by these prior researches, this work aims to study the greybody bound for the static black hole in $f(R)$ gravity.

In GR, the limitation of the cosmological constant includes the discrepancies of flat galactic rotation curves, accelerated expansion of the universe, anti-lensing, the observed anisotropies on the cosmic microwave background radiation (CMB), and the problem of coincidence \cite{c1,c2,c3,c4,c5,c6}. Many people suspect that these phenomena arise from the dark side of our universe which so far has not been addressed properly. For instance, introducing the simple cosmological constant term, as a nonzero vacuum energy, to the Einstein field equations can regenerate the acceleration of the universe. However, the argument for the small value of the cosmological constant
has yet been inconspicuous. A partial solution to this problem is the consideration of varying dark energy model in which the density of dark energy can find the density of matter up to the present time from the early Universe. To analyze the unresolved cosmological puzzles such as the late time acceleration expansion, we should focus on the modified theories of gravity to impersonate the effects of dark energy and dark matter and to get an effective time-varying equation of state. In such models, in accordance with the indispensability and in order to explain the dynamics of the universe in galactic, astrophysical, or cosmic scales, the Einstein-Hilbert action is modified or extended. People have been making an effort, since the last decades, to solve these problems with the support of modified theories of gravity. In particular, in Refs. \cite{c7,c7a,c7b}, authors have reported some intuitive extensions of GR after the replacement of the Einstein-Hilbert action with a generic form of $f(R)$ theory. As a consequence, the $f(R)$ theories of gravity, during the last decades, have been of great interest and have been examined to check the consistency (see e.g. Refs. \cite{v1,v2,v3,v4,v5,v6,v7} and the reviews \cite{v8,v9}). However, as in GR, our interest lies in the black hole solutions that are extracted in view of $f(R)$ gravity. In view of Palatini formalism, among the two approaches namely metric and Palatini formalism, the solution is the Schwarzschild–de Sitter metric along with effective cosmological constant, which appears to endure from inconsistency with primary tests in GR as the cosmological constant has no crucial role in solar system scales \cite{v10}. In order to avoid this problem, one can manipulate the action such that the effectiveness of the cosmological constant becomes insignificant in the solar system scales and crucial in the cosmological scales \cite{v11,v12}. Accordingly, an appropriate $f(R)$ action model has been introduced in Refs. \cite{c10,c12} which is
compatible with both galactic and cosmological scales. Moreover, in Ref. \cite{f1}, this model has been elucidated by virtue of a generic function in the gravitational action, as a result, it is compatible with the all mentioned scales such as solar system scale, galactic scale as well as cosmological scale. In addition to this, the authors of Ref. \cite{f1}, in the context of $f(R)$ gravity, have also developed a black hole solution which is static and spherically symmetric and this solution is also our interest in this work, regarding the weak deflection angle, Hawking radiation, gerybody factor, and its shadow.

Moreover, the complementary cosmological tests of this $f(R)$ gravity model have been conducted in Ref. \cite{f1} by utilizing the observational data of the Pioneer anomalies in the solar system, the rotation curve of the galaxies, and supernova type Ia gold sample (SN Ia) in the cosmological scales to put a constraint on the parameters of this model. It is remarkable that the supernova type Ia gold sample data gives compatible results with the other observations. Apart from this, S. Asgari et al. \cite{mond} have considered the $f(R)$ gravity model in galactic scale which in turn gives solutions to be consistent with spherically symmetric space-time. However, the results of this consideration, in addition to the verification of flat rotation curves in spiral galaxies, is in complete agreement with Tully-Fisher (TF) relation and is compatible with MOdified Newtonian Dynamics (MOND) theory.

It is expected that, GR heeds gravity with high exactitude when the curvature is small \cite{c8}, however, there does not subsist any such evidence for very broad/large values of curvature. In this connection, black holes become the norm places to look for modified GR \cite{c9}. In $f(R)$ modified gravity models, at large distances, the geometry of the space-time should be different from that of standard GR. Hence, we need to introduce a method that could observationally discriminate and test the possible deviations from GR. The study of weak deflection angle, greybody factor, and shadow cast by the static spherically symmetric $f(R)$ black hole could be one of the possible methods. In 2016, S. Soroushfar et al. \cite{i1} studied the thermodynamical behavior, stability conditions, and phase transition of static, static-charged, and rotating-charged black holes in $f(R)$ gravity. Also, they extend their study to various thermodynamic geometry methods such as Ruppeiner, Weinhold, and Geometrothermodynamics (GTD). Later (2019), the study of thermal corrections and phase transition considering the static black hole in $f(R)$ gravity has been carried out by M. Rostami et al. \cite{hw}. The investigation of accretion disc for a static spherically symmetric black hole in $f(R)$ gravity can be found in Ref. \cite{i2}. The geodesic study \cite{i3} and study of perturbed thermodynamics and thermodynamic geometry \cite{g1} of a static charged black hole spacetime in the framework of $f(R)$ gravity has been established. A brief study on the features of static and spherically symmetric solutions with a horizon in $f(R)$ theories can be found in Ref. \cite{BS}. Recently (2022), the study of the trajectory of massive particles in the vicinity of a static black hole in $f(R)$ gravity was accomplished by S. Mandal et al. (see Ref. \cite{f2a}). In this continuity, here in the present study, we intend to investigate the weak deflection angle by utilizing the GBT in a non-plasma medium, Hawking radiation, and the bound on greybody factor for a static black hole in the environments of $f(R)$ gravity. We also study the shadow cast by this black hole. Our study also focuses on how the cosmological constant and non-zero real constant (introduced in static spherically symmetric $f(R)$ black hole) affect the deflection angle, Hawking temperature, greybody bound, and silhouette cast by static spherically symmetric $f(R)$ black hole.

This paper is presented in nine parts. In section \ref{sec3}, we give a brief outline of a static black hole solution in $f(R)$ gravity, and in the framework of weak gravitational lensing, we derive corresponding optical metric and Gaussian curvature. In section \ref{sec4}, we study GBT in detail and derive the deflection angle of such a black hole in the presence of a non-plasma medium. In section \ref{sec4a}, by doing a graphical analysis, we focus mainly on how the calculated deflection angle depends on various parameters that govern our considered black hole. We explore the Hawking temperature based on GBT and a graphical interpretation is also accomplished to see the dependency of Hawking temperature on various parameters in section \ref{sec5}. In section \ref{sec6}, we estimated the potential and rigorous bounds of the greybody factor (transmission probability) of the static $f(R)$ black hole. The graphical seasoning of the potential and the corresponding bound on the greybody factor are addressed in section \ref{sec7}. We have a discussion on null geodesics for the static black hole in the environments of $f(R)$ model in section \ref{sec8}. In section \ref{sec9}, we present deliberation related to black hole shadow in the framework of a non-plasma medium. The shape of the silhouette is executed with the employment of the geodesic equations of a test particle near our considered black hole. In this section, we also addressed a graphical analysis to observe the dependency of shadow radius on various parameters. Also, we did a comparative plot for the angular diameter of the static $f(R)$ black hole with observed angular diameter of M$87^{\star}$ and Sgr $A^{\star}$ by the EHT and constrain the parameter $\beta$ and $\Lambda$ of the spacetime. Additionally, the correspondence between the shadow radius and the real part of QNMs in \textit{eikonal regime} is discussed. Finally, Section \ref{sec10} is devoted to concluding the results obtained from our analysis and to making final remarks. We consider the metric signature  ($-, +, +, +$) throughout the paper.

\section{Static black hole in  \texorpdfstring{$f(R)$}{Lg} gravity}\label{sec3}
In this section, we introduce the metric structure of the static spherically symmetric black hole in the context of $f(R)$ gravity\footnote{In the context of GR, $f(R)$ gravity is a class of modified gravitational theories in which the Einstein-Hilbert action is generalized to comprehend a non-linear function of the Ricci scalar curvature, $R$, instead of the usual linear term. This type of theory was introduced as an alternative to dark matter, which is thought to be indispensable in explaining certain observational phenomena in the universe, such as the rotation of galaxies as well as the formation of large-scale structures. In $f(R)$ gravity, the excessive terms in the action modify the equations of motion for gravity and can give rise to additional forces that can mimic the effects of dark matter without the need for additional matter.}. To embark on our investigation in 4-dimensional spacetime recalling $f(R)$ gravity, the generic action based on the Ricci scalar is articulated as \cite{f1}
\begin{equation}\label{32}
\mathcal{I}=\frac{1}{2}\int d^4x\sqrt{-g}f(R)+\mathcal{I}_{m}
\end{equation}
where $f(R)$ is a function of the Ricci scalar ($R$), and the matter part of the above action is denoted by $\mathcal{I}_{m}$. Here, we set the natural unit $G=c=\hbar=\kappa=1$. Recapitulating the variation principle of the action $\mathcal{I}$ presented in (\ref{32}) with respect to metric leading to the following field equations :
\begin{equation}\label{33}
f^{\prime}(R)R_{\mu\nu}-\frac{1}{2}g_{\mu\nu}f(R)-(\nabla_{\mu}\nabla_{\nu}-g_{\mu\nu}\square)f^{\prime}(R)=T_{\mu\nu}
\end{equation}
with $\square = \nabla_{\alpha}\nabla^{\alpha}$, $f^{\prime} = \frac{df(R)}{dR}$, and $T_{\mu\nu}$ is the energy -momentum tensor. The proposed vacuum solution to the field equation \ref{33} for a 4-dimensional static spherically symmetric black hole in f(R) gravity is given by the line element \cite{f1}
\begin{equation}\label{34}
ds^2=-B(r)dt^2+\frac{dr^2}{B(r)}+r^2(d\theta^2+sin^2\theta d\phi^2),
\end{equation}
in the conventional Schwarzschild coordinates $x^{\mu}=(t,r,\theta,\phi)$. The model chosen for $f(R)$ gravity is given by \cite{f1}
\begin{equation}\label{35}
f(R)=R+\Lambda+\frac{R+\Lambda}{\frac{R}{R_{0}}+\frac{2}{\alpha}}ln\frac{R+\Lambda}{R_{c}}
\end{equation}
where $R_{0} = \frac{6\alpha^2}{d^2}$ in which $\alpha$ and $d$, represent free parameters of the action, $\Lambda$ denotes the cosmological constant, and $R_{c}$ is a constant extracting from performing the integration of the action \ref{32}. Here, $\alpha$ has no dimension whereas $[d]=m$.
The cosmological constant takes the value of $|\Lambda| \le 10^{-52} m^{-2}$ as suggested in Refs. \cite{f2,f2a}. A reader can see the detailed derivation of the metric solution \cite{f1,fr}. In the sequel, the metric function (or, lapse function) designed up to the first order in the free parameters of the mentioned action \ref{32} becomes 
\begin{equation}\label{36}
B(r)=1-\frac{2M}{r}+\beta r-\frac{1}{3}\Lambda r^2
\end{equation}
where $M$ is a constant pertained with the mass of the black hole, $\beta=\frac{\alpha}{d}\ge0$ denotes a real constant \cite{f1,f2,f2a,f3} which is related to the dark-matter. In this model, $d$ is a scale factor that is around 10 kpc, and the dimensionless parameter $\alpha\simeq 10^{-6}$ which describes a flat rotation curve of stars in a typical spiral galaxy. However, one should note that non-zero cosmological constant has intricate consequences in various perspectives of static and spherically symmetric black holes as documented in Refs. \cite{f4, f5}. The conception of the static radius is significant in an extending universe subservient by a cosmological constant as it can personate a natural boundary of gravitationally bound systems, as addressed in various situations \cite{f6, f7, f8, f9, f10, f11}. However, in the scenario when $\beta=0$, one can impeccably retrieve the space-time of the Schwarzschild AdS black hole. It is noteworthy to mention that performing the limit $\beta=\Lambda=0$ in \ref{36} leads to the reduction of the well-known Schwarzschild solution.

It is worth mentioning that the static spherically symmetric vacuum solution to the first order in free parameters in $f(R)$ gravity describes an uncharged black hole which corroborates with the Mannheim-Kazanas (MK) vacuum solution to the fourth-order Weyl conformal gravity \cite{frr}. In the MK vacuum solution, the linear term $\beta r$ plays the character of an extra potential recoup for the galactic rotation curves (flat). In the same manner, for a typical galaxy, the model written in Eq. \ref{36} raises that low values of $\alpha$ can give the flat galactic rotation curves. However, for strong curvature, the model mentioned in \ref{35} is reduced to
\begin{equation}\label{36a}
f(R)=R+R_{0}\ln\frac{R}{R_{c}},
\end{equation}
in which $R>>\Lambda$ and $\frac{R}{R_{0}}>>\frac{2}{\alpha}$ which is affiliated to the case of stellar black holes. Nevertheless, in the
cosmological scales (where $R\simeq R_{0} \simeq \Lambda$ and $\alpha<<1$), the model shortened to $f(R)=R+\Lambda$ which can respond to the Einstein-Hilbert action along with a cosmological constant that portrayed the late time accelerated expansion of the universe. Thus, the low-valued non-negative free parameter $\beta$ given in Eq. \ref{36} can veil both the small ($R>>\Lambda$) and large-scale ($R\approx\Lambda$) phenomena in the universe.

The horizon structure of this black hole has been presented in Refs. \cite{i2,f2a}. Here, it is worth mentioning that, due to small values of $\Lambda$ and positive values of $\beta$ (wide range), for the sake of simplicity, all the graphs are scaled with $\Lambda$ ranging $1\le \Lambda<50$ and $\beta$ ranging $0<\beta<60$ whereas $M$ taking values in the range $1\le M<40$. This consideration is to make the graphical analysis more simpler and it has no other physical meaning.

\subsection{Optical metric and Gaussian curvature in weak gravitational lensing}
This section focuses on null geodesics deflected by the above-mentioned black hole. It is well-established that light satisfies the null geodesic (i.e. $ds^2=0$). This null geodesic is meticulously designed to define the optical metric that delineates Riemannian geometry followed by light. Now, imposing the null condition unveils the following optical metric (in new coordinate system $\tilde{r}$):
\begin{equation}\label{37}
dt^2=\bar{g}_{ij} dx^i dx^j= d\tilde{r}^2+B(\tilde{r})^2d\phi^2
\end{equation}
where  \begin{equation}\label{38}
d\tilde{r}=\frac{dr}{(1-\frac{2M}{r}+\beta r-\frac{1}{3}\Lambda r^2)},
\end{equation}
\begin{equation}\label{39}
B(\tilde{r})=\frac{r}{\sqrt{1-\frac{2M}{r}+\beta r-\frac{1}{3}\Lambda r^2}}.
\end{equation}
It is remarkable that the previous system $(i, j)$ is recast into a new system $(r, \phi)$ giving the determinant value as $det(\bar{g}_{ij})=\frac{r^2}{B(\tilde{r})^3}$. It is indisputable that the equatorial plane in the optical metric offers a prospective surface of revolution. The non-vanishing Christoffel symbols affiliated to metric (\ref{37}) are computed as 
\begin{eqnarray}
\Gamma^{\tilde{r}}_{\phi\phi}&=&\frac{r(rB^\prime(\tilde{r})-2B(\tilde{r}))}{2},\\
\Gamma^{\phi}_{\tilde{r}\phi}&=&\frac{2B(\tilde{r})-rB^\prime(\tilde{r})}{2},\\\Gamma^{\tilde{r}}_{\tilde{r}\tilde{r}}&=&-\frac{B^\prime(\tilde{r})}{B(\tilde{r})}.
\end{eqnarray} 
Here, prime (dash) replicates the derivative with respect to $r$. The above Christoffel symbols offer a unique Riemann tensor for optical curvature that takes non-zero value: $R_{\tilde{r}\phi\tilde{r}\phi}$=-$kB^2(\tilde{r})$.
A particularly noteworthy connection route that manifests the naive interplay between the Gaussian optical curvature $\mathcal{K}$ and the Ricci Scalar $R_{icciScalar}$ as
\begin{equation}\label{41}
\mathcal{K}=\frac{R_{icciScalar}}{2}=-\frac{1}{B(\tilde{r})}\left[\frac{d r}{d \tilde{r}} \frac{d}{d r}\left(\frac{d r}{d \tilde{r}}\right) \frac{d B}{d r}+\frac{d^{2} B}{d r^{2}}\left(\frac{d r}{d\tilde{r}}\right)^{2}\right].
\end{equation}
By reminiscing Eqs. (\ref{38}) and (\ref{39}), in the sequel, Gaussian optical curvature can be written in the following  explicit form:
\begin{eqnarray}\label{42}
\mathcal{K}&=&-\Big(\frac{1}{3}\Lambda+\frac{\beta^2}{4}\Big)+\frac{2M\Lambda}{r}-\frac{3M\beta}{r^2}-\frac{2M}{r^3}+\frac{3M^2}{r^4}\nonumber\\
 &+& O(M^3,\Lambda^2,\beta^3).
\end{eqnarray}
Here, it is evident that the Gaussian optical curvature has a dependency on various parameters such as mass $M$, cosmological constant $\Lambda$, and real constant $\beta$.

\section{Deflection angle of a static black hole in \texorpdfstring{$f(R)$}{Lg} gravity in non-plasma medium}\label{sec4}
This section aims to elucidate the process of calculating the deflection angle, in weak field approximation, of a black hole resembling the static black hole within the framework of $f(R)$ gravity and this study is in the non-plasma medium by employing the GBT. The GBT, which plays a crucial role in connecting the (intrinsic) geometry of metric with the underlying topology in the region $\mathcal{V}_{\cal {R}}$ with boundary $\partial \mathcal{V}_{\cal {R}}$, as elegantly enunciated below \cite{f12,f12a}
\begin{equation}\label{43}
\iint_{\mathcal{V}_{\cal R}} \mathcal{K} d S+\oint_{\partial \mathcal{V}_{\cal R}} k d t+\sum_{z} \alpha_{z}=2 \pi  {\Xi}\left(\mathcal{V}_{\cal {R}}\right),
 \end{equation}
in which  $\mathcal{V}_{\cal R}\subset S$ is a regular domain of a simple two-dimensional surface $S$ having closed, regular, and positive oriented boundary $\partial \mathcal{V}_{\cal R}$. Here, $k$ denotes the geodesic curvature of $\partial \mathcal{V}_{\cal R}$ define as $k=\bar{g}\left(\nabla_{\dot{\gamma}} \dot{\gamma}, \ddot{\gamma}\right)$, in which $\bar{g}(\dot{\gamma}, \dot{\gamma})=1$ and  $\ddot{\gamma}$ indicating unit acceleration vector. At the $z^{\mbox{th}}$ vertex$, \alpha_{z}$ represents the exterior angle. Also, here in Eq. \ref{43}, $\Xi\left(\mathcal{V}_{\cal {R}}\right)$ refers to the Euler characteristic number. If we consider the limit $\cal{R} \rightarrow \infty$, we can acquire the jump angle with value $\pi/2$ so that $\theta_{0}+\theta_{S}=\pi$ \cite{f12b} and Euler the characteristic number takes a unit value. Consequently, one can expect the following result:
\begin{equation}\label{43a}
\iint_{\mathcal{V}_{\cal R}} \mathcal{K} d S+\oint_{\partial \mathcal{V}_{\cal R}} k d t+\alpha_{z}=2 \pi  {\Xi}\left(\mathcal{V}_{\cal {R}}\right),
\end{equation}
where the total angle of jumps is denoted by $\alpha_{z}=\pi$, and since $S\rightarrow\infty$, the geodesic curvature can be expressed as $k\left(E_{\cal R}\right)= |\nabla_{\dot{E}_{\cal R}} \dot{E}_{\cal R}|$. The radial component of geodesic curvature can be acquired as \cite{f12,f12a}
\begin{equation}\label{44}
\left(\nabla_{\dot{E}_{\cal R}} \dot{E}_{\cal R}\right)^{r}=\dot{E}_{\cal R}^{\phi} \partial_{\varphi} \dot{E}_{\cal R}^{r}+\Gamma_{\varphi \varphi}^{\tilde{r}}\left(\dot{E}_{\cal R}^{\phi}\right)^{2}.
\end{equation}
At large ${\cal R}$, we have $E_{\cal R}:=r(\phi)={\cal R}=$ constant resulting in   $\left(\dot{E}_{R}^{\phi}\right)^{2}=\frac{1}{f^{2}(\tilde{r})}$.
Recalling $\Gamma_{\varphi \phi}^{\tilde{r}}=\frac{r\left(r f^{\prime}(\tilde{r})-2 f(\tilde{r})\right.}{2}$, geodesic curvature obtained as \cite{f12}
 \begin{equation}\label{45}
 \left(\nabla_{\dot{E}_{\cal R}^{r}} \dot{E}_{{\cal R}}^{r}\right)^{r} \rightarrow \frac{1}{\cal R}.
 \end{equation}
This indicates that $k(E_{\cal R})\rightarrow \frac{1}{\cal R}$. We can have $d t={\cal R} d \phi$ by employing optical metric (\ref{37}). In the sequel, one might get
 \begin{equation}\label{46}
 k(E_{\cal R})dt=\lim_{{\cal R}\to\infty}[k(E_{\cal R})dt]
         =\lim_{{\cal R}\to\infty}\left[\frac{1}{2\sqrt{\bar{g}_{rr}  \bar{g}_{\phi\phi}}}(\frac{\partial\bar{g}_{\phi\phi}}{\partial r})\right]d\phi
         =d\phi.
 \end{equation}
Considering all the above results into account, the GBT becomes 
 \begin{equation}\label{47}
 \iint_{\mathcal{V}_{{\cal R}}} \mathcal{K} d S+\oint_{\partial \mathcal{V}_{{\cal R}}} k d t\stackrel{{\cal R}\rightarrow \infty} {=}  \iint_{S_{\infty}} \mathcal{K} d S+\int_{0}^{\pi+\tilde{\delta}} d \phi.
 \end{equation}
  
In the framework of weak deflection limit, the light ray for the zeroth order apprehends by a straight line approximation as $r(t)=\frac{b}{sin\varphi}$, where $b$ is related to the impact parameter. By utilizing this formula, the deflection angle can be calculated as
\begin{equation}\label{48}
\tilde{\delta}=-\int_{0}^{\pi} \int_{b / \sin \varphi}^{\infty} \mathcal{K} dS=-\int_{0}^{\pi} \int_{b / \sin \varphi}^{\infty} \mathcal{K} \sqrt{\operatorname{det} \bar{g}} d r d \phi =-\int_{0}^{\pi} \int_{b / \sin \varphi}^{\infty} \mathcal{K} \frac{r}{f(r)^\frac{3}{2}} d r d \phi
\end{equation}

In the environments of the non-plasma medium, for the mentioned metric function of a static black hole in $f(R)$ gravity, the deflection angle is simplified to
\begin{eqnarray}\label{49}
\tilde{\delta}&=&-\int_{0}^{\pi} \int_{b / \sin \varphi}^{\infty}\left[\frac{15M \beta^2}{4}-\Big(\frac{1}{3}\Lambda+\frac{9}{2}M\Lambda\beta+\frac{\beta^2}{4}\Big)r+\frac{1}{2}\Lambda\beta r^2+\frac{15M^2\Lambda}{2r}+\frac{1}{r^2}\Big(-\frac{27}{2}M^2\beta-2M\Big)\right.
\nonumber\\
&-&\left. \frac{3M^2}{r^3}+O(M^3,\Lambda^2,\beta^2) \right] dr d \phi.
\end{eqnarray}
\begin{eqnarray}\label{50}
\tilde{\delta}&=&\frac{15M\beta^2b}{4}\int_{0}^{\pi}\left.\csc\phi\ d \phi-\Big(\frac{1}{6}\Lambda+\frac{9}{4}M\Lambda\beta+\frac{\beta^2}{8}\Big)b^2 \int_{0}^{\pi}\csc^2\phi\ d \phi+
\frac{\Lambda\beta b^3}{6}\int_{0}^{\pi}\csc^3\phi \ d \phi\right.\nonumber\\&+&\frac{15M^2\Lambda}{2}\Big\{\pi \log(b)-\left.\int_{0}^{\pi}\log(\sin\phi)\ d\phi\Big\}+ \frac{(27M^2\beta+4M)}{b}\int_{0}^{\pi}\sin\phi\ d \phi\right.\nonumber\\&-&\frac{3M^2}{2b^2}\left.\int_{0}^{\pi}\sin^2\phi\ d \phi\right.
\end{eqnarray}
Since the integral of $csc\phi$ does not converge on $\{0,\pi\}$ and it is notified from the above Eq. (\ref{50}) that the first three terms do not contribute to the deflection angle. Therefore, the explicit expression for the deflection angle for a static black hole in $f(R)$ gravity in the weak limit becomes
\begin{equation}\label{51}
\tilde{\delta}=\frac{15M^2\Lambda}{2}\Big\{\pi \log(b)+\pi \log(2)\Big\}+\frac{1}{b}\Big(27M^2\beta+4M\Big)-\frac{3M^2\pi}{4b^2}+O(M^3,\Lambda^2,\beta^2) 
 \end{equation}
Here, it is evident that the deflection angle of the static black hole in $f(R)$ gravity depends on the various parameters such as impact parameter $b$, the mass of black hole $M$, real constant $\beta$ and cosmological constant $\Lambda$. So, it is expected that there may be a significant change in $\tilde{\delta}$ because of the different parameters that govern the black hole. It is remarkable that by overriding the term that contains $\beta$ in the above calculated deflection angle, one can attain the deflection angle of Schwarzschild Ads black hole. Nevertheless, one can achieve the deflection angle for Schwarzschild black hole up to the second order of mass when both $\Lambda=\beta=0$.

 \begin{figure}[ht]
\begin{center} 
 $\begin{array}{cccc}
\subfigure[]{\includegraphics[width=0.5\linewidth]{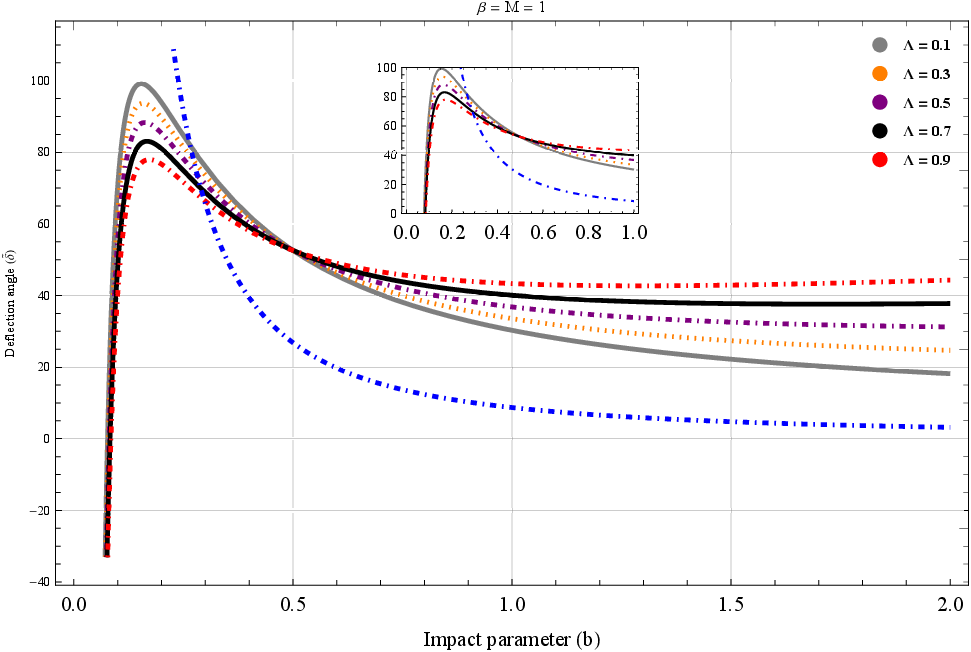}
\label{1a}}
\subfigure[]{\includegraphics[width=0.5\linewidth]{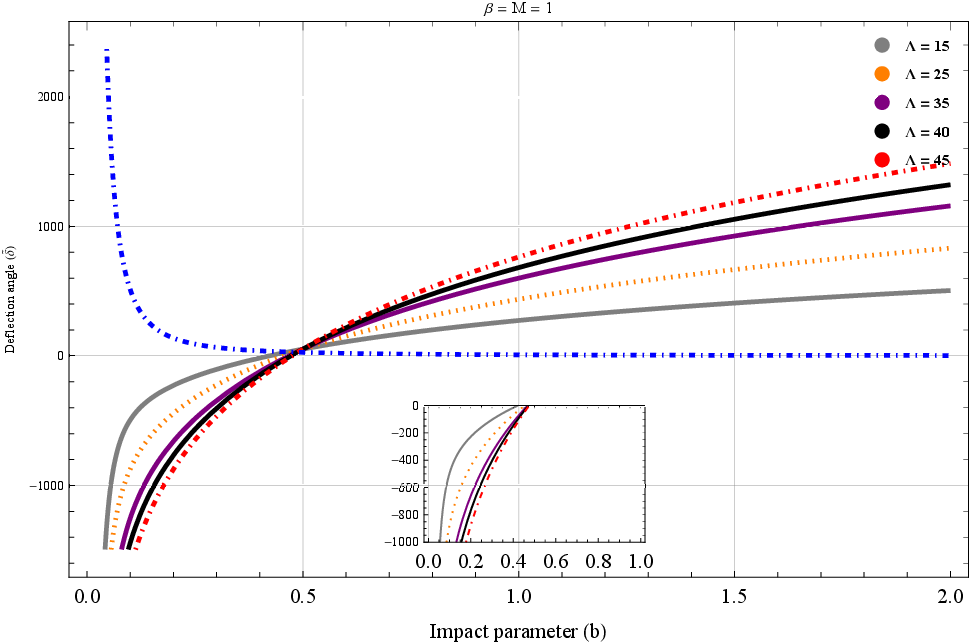}\label{1b}}\\
\subfigure[]{\includegraphics[width=0.5\linewidth]{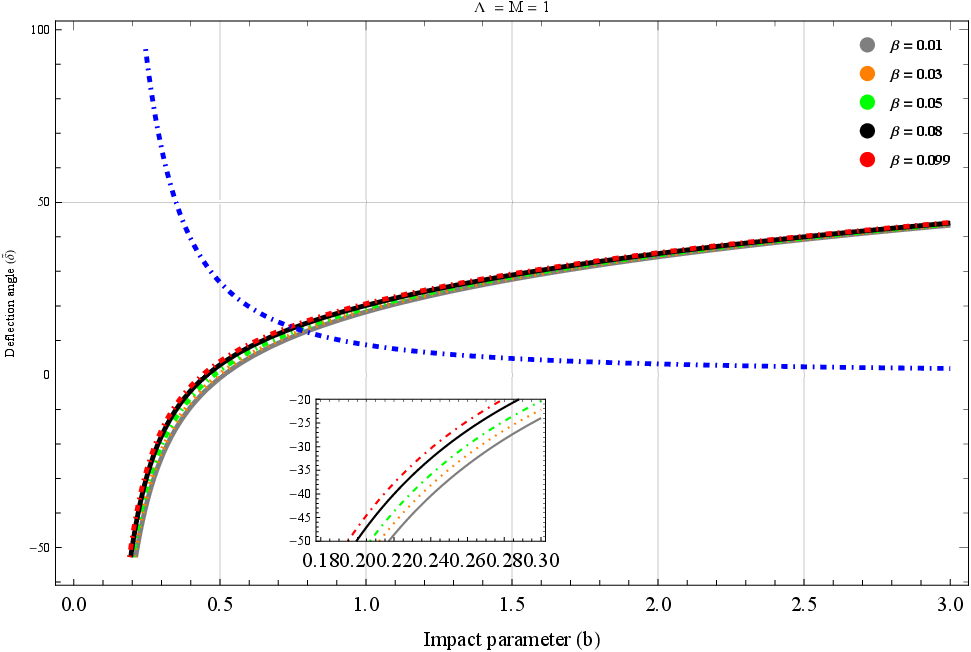}\label{1c}}
\subfigure[]{\includegraphics[width=0.5\linewidth]{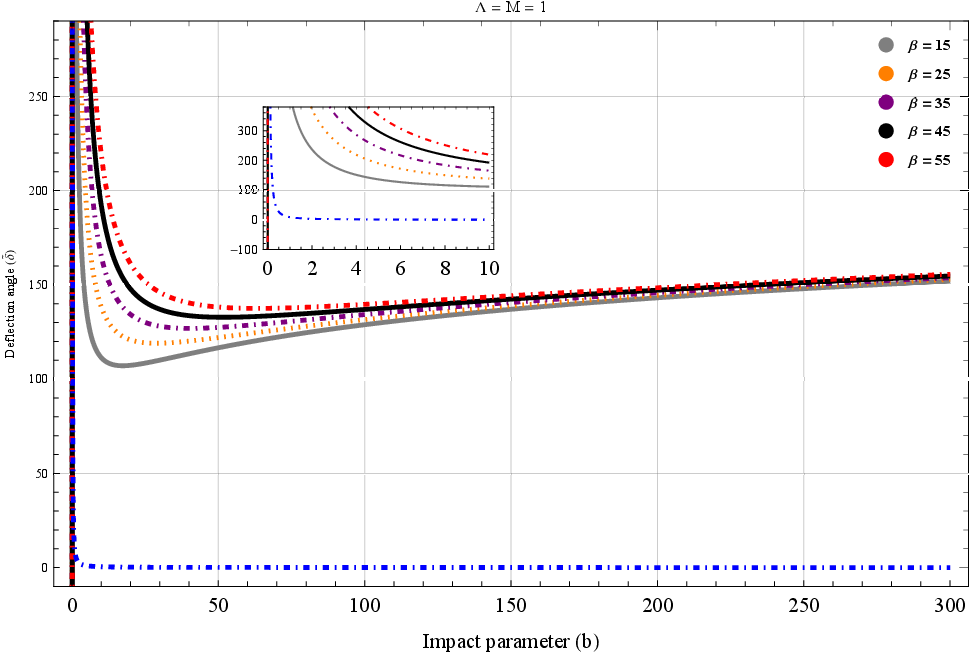} \label{1d}}\\
\subfigure[]{\includegraphics[width=0.5\linewidth]{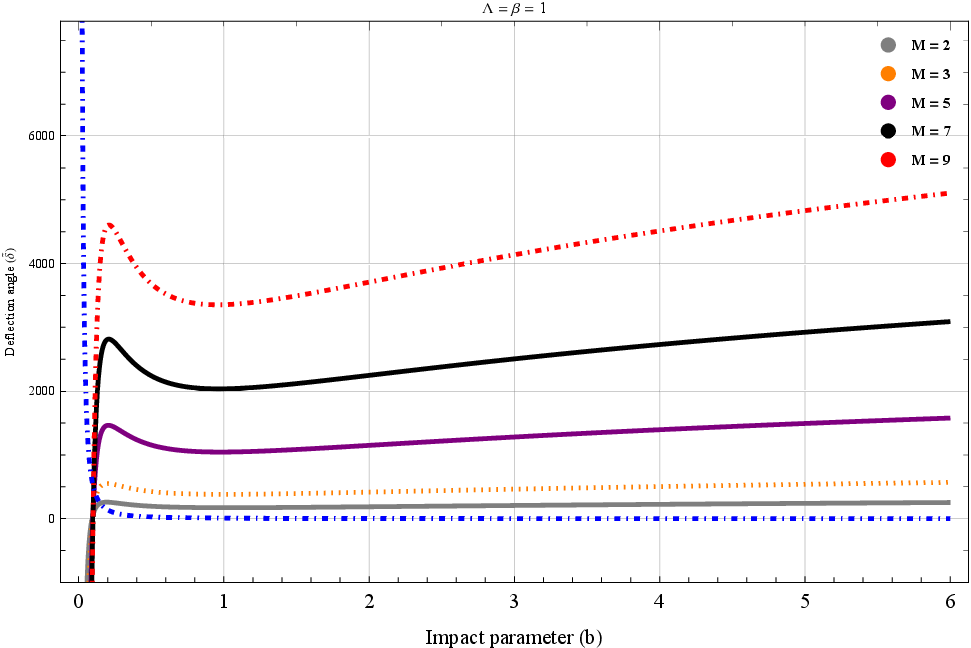} \label{1e}}
\subfigure[]{\includegraphics[width=0.5\linewidth]{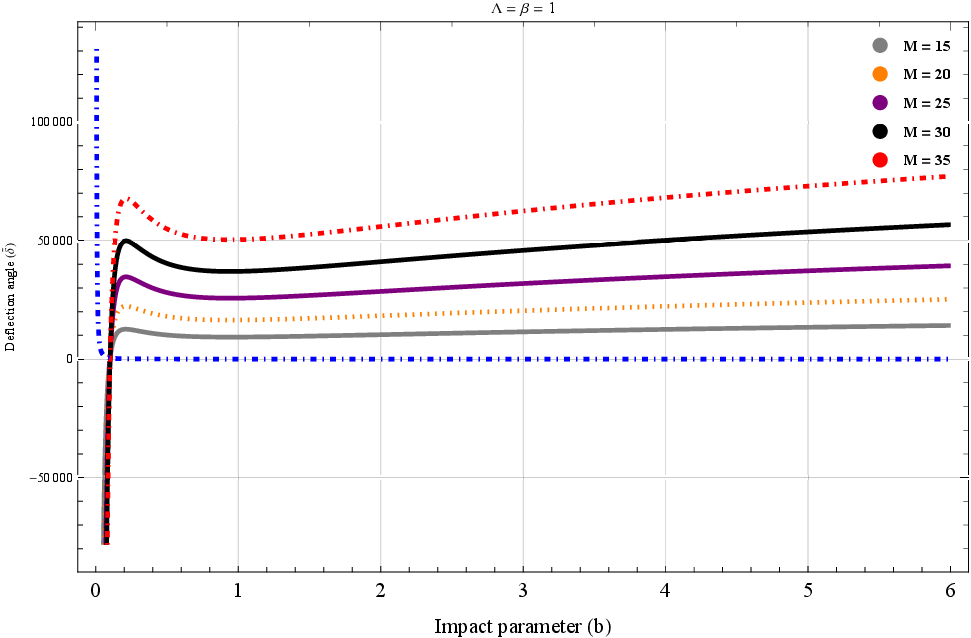} \label{1f}}
\end{array}$
\end{center}
\caption{The graph of deflection angle ($\tilde{\delta}$) with respect to impact parameter ($b$). In (a) and (b): for varying $\Lambda$ but fixed $\beta = M = 1$. In (c) and (d): for varying $\beta$ but fixed $\Lambda = M = 1$. In (e) and (f): for varying $M$ and for $\Lambda = \beta= 1$. Inset shows the behavior of $\tilde{\delta}$ for lower $b$. }
\label{fig1}
\end{figure}
 
\section{Graphical analysis of deflection angle for non-plasma medium}\label{sec4a}
In this section, we study the behavior of deflection angle and their dependence on various parameters. 
\subsection{Effect of impact parameter (\texorpdfstring{$b$}{Lg}) on deflection angle (\texorpdfstring{$\tilde{\delta}$}{Lg})}
We plot the behavior of the deflection angle and its dependence on the impact parameter ($b$) for changing $\Lambda$, $\beta$, and $M$ in Fig.\ref{fig1}. Here, from Fig.\ref{1a}, it is clear that the lower (positive) value of the cosmological constant enforces the deflection angle to start from a negative region and attains its maximum values by showing a peak in the positive region for lower $b$ and then increases softly with $b$ in positive region after showing declination behavior. In contrast, for a higher (positive) valued cosmological constant (Fig.\ref{1b}, $\tilde{\delta}$ begins from the negative region by illustrating the increasing nature for smaller $b$ while for higher $b$, it goes to the positive region by retaining increasing tendency. Moreover, the deflection angle, for both cases, decreases (increases) with increasing cosmological constant for small (large) $b$. In Fig.\ref{1c}, the deflection angle shows an increasing tendency with impact parameter $b$ for small values of $\beta$ and remains positively valued in the higher region of $b$. Here, $\tilde{\delta}$ always increases for increasing $\beta$. However, in Fig.\ref{1d}, for large $\beta$, $\tilde{\delta}$ become asymptotical for small $b$, afterthat it shows same trend as that of Fig.\ref{1a}. Here also $\tilde{\delta}$ increases when $\beta$ values become higher. However, analysing Figs.\ref{1e} and \ref{1f}, we verify that for both small and large $M$, initially $\tilde{\delta}$ increases abruptly for small $b$ and it finally deliberates increasing demeanor for large $b$ after performing certain decline nature. Here, $\tilde{\delta}$ takes only positive values and it increases with increasing $M$.

\begin{figure}[ht]
\begin{center} 
$\begin{array}{cccc}
\subfigure[]{\includegraphics[width=0.5\linewidth]{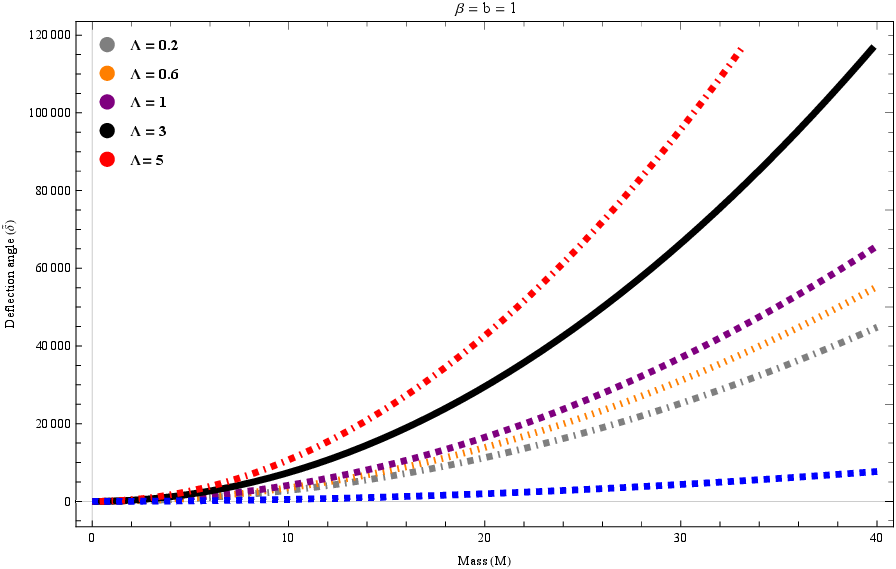}
\label{2a}}
\subfigure[]{\includegraphics[width=0.5\linewidth]{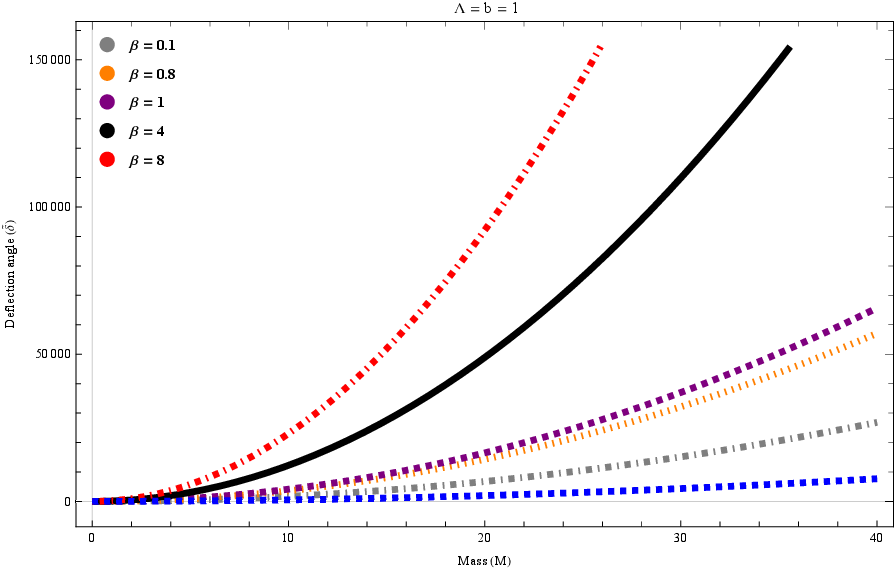}\label{2b}}\\
\subfigure[]{\includegraphics[width=0.5\linewidth]{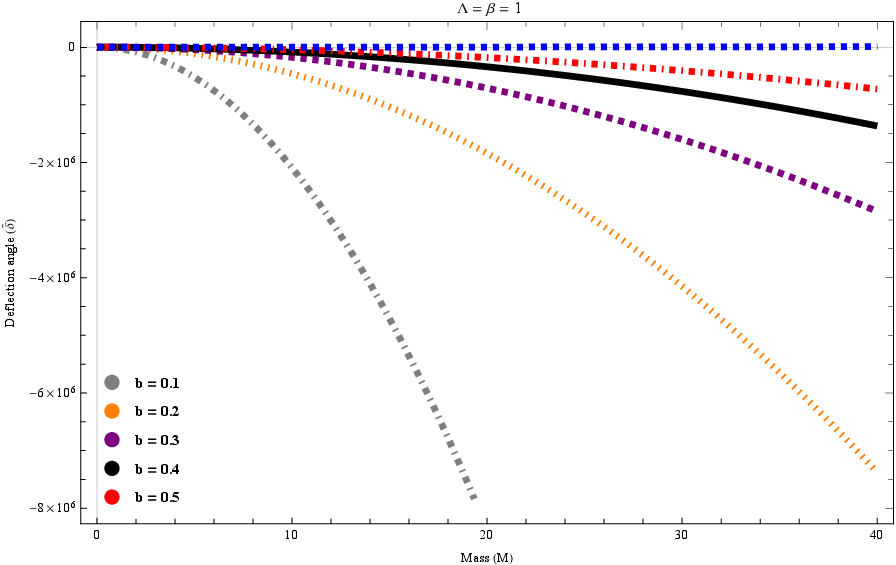}
\label{2c}}
\subfigure[]{\includegraphics[width=0.5\linewidth]{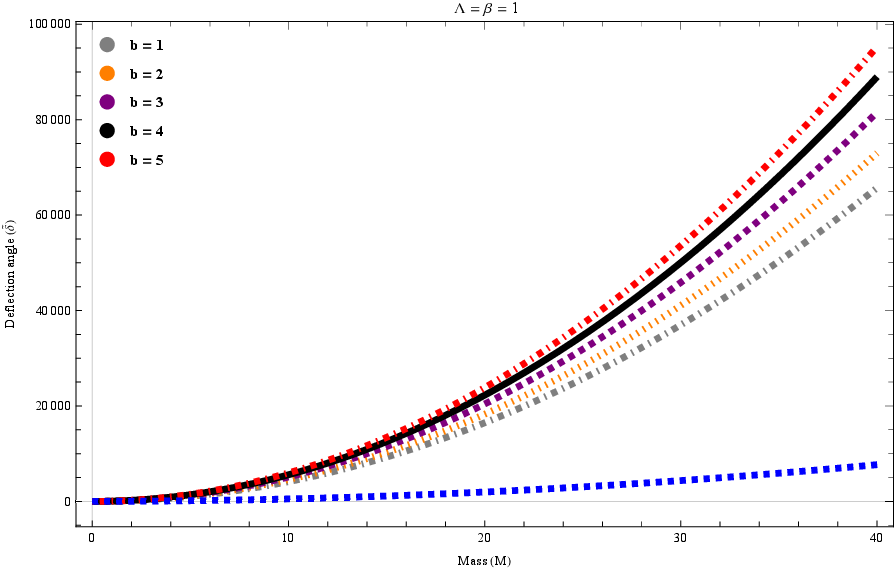}\label{2d}}  
\end{array}$
 \end{center}
\caption{The behavior of deflection angle ($\tilde{\delta}$) with respect to mass $M$ by varying  $\Lambda$ with fixed $\beta = b= 1$ [(a)], by varying $\beta$ with fixed $\Lambda=b=1$ [(b)] and for changing $b$ with fixed $\Lambda = \beta = 1$ [(c)-(d)].}
\label{fig2}
\end{figure}
\subsection{Effect of mass (\texorpdfstring{$M$}{Lg}) on deflection angle (\texorpdfstring{$\tilde{\delta}$}{Lg})}
The variation of deflection angle $\tilde{\delta}$ with $M$ for different $\Lambda$, $\beta$, and $b$ is depicted in Fig.\ref{fig2}. We notice that the deflection angle is an increasing function of $M$ for all values of $\Lambda$ and $\beta$ and it remains positively valued. In contrast, $\tilde{\delta}$ decreases with $M$ for small values of $b$ and small values of $\beta$ and takes negative values only. Moreover, for the larger value of $b$, the deflection angle performs increasing behavior.

\begin{figure}[ht]
\begin{center} 
$\begin{array}{cccc}
\subfigure[]{\includegraphics[width=0.5\linewidth]{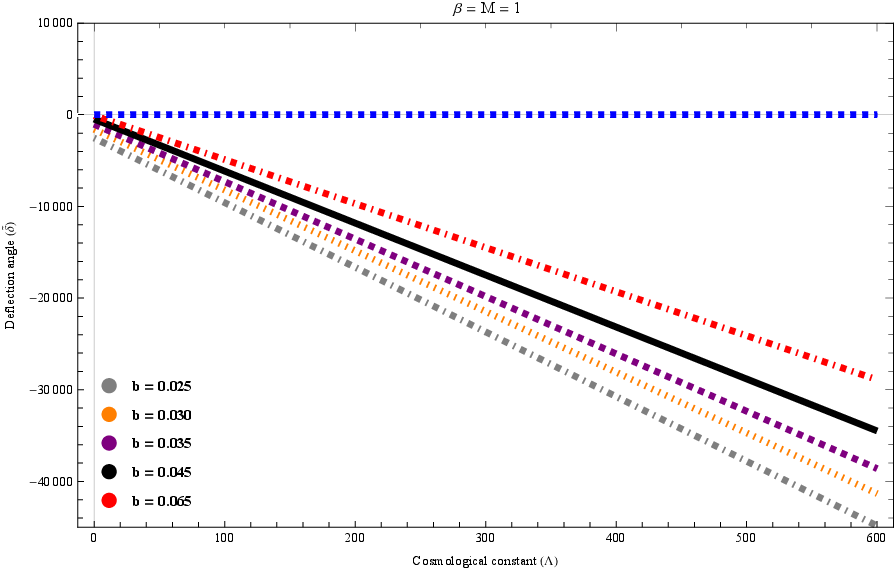}\label{3a}}
\subfigure[]{\includegraphics[width=0.5\linewidth]{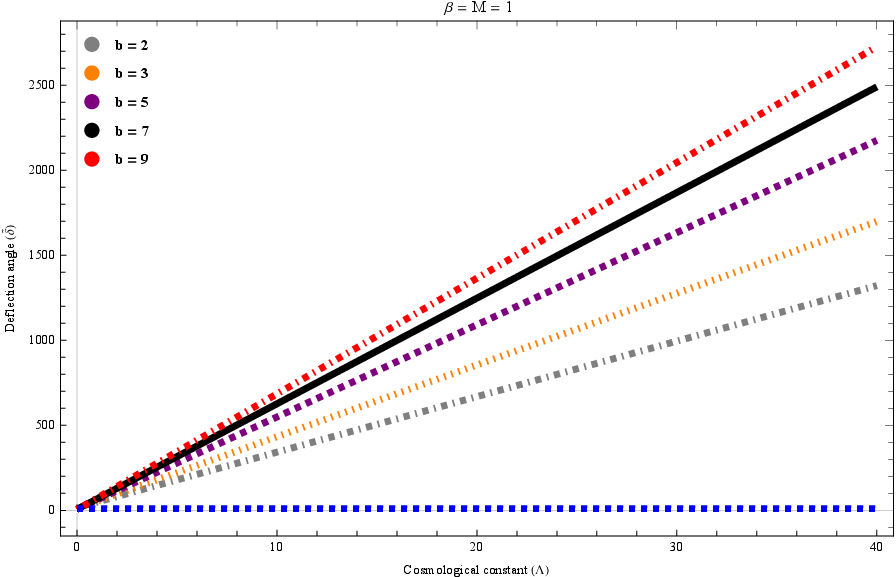}\label{3b}}\\
\subfigure[]{\includegraphics[width=0.5\linewidth]{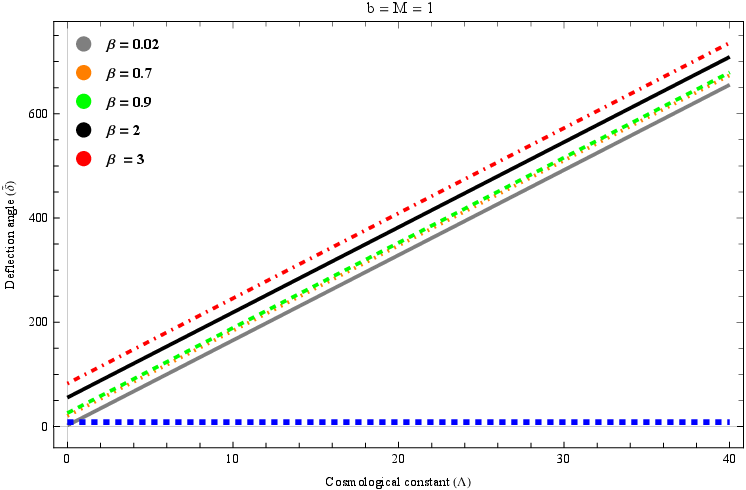}\label{3c}}
\subfigure[]{\includegraphics[width=0.5\linewidth]{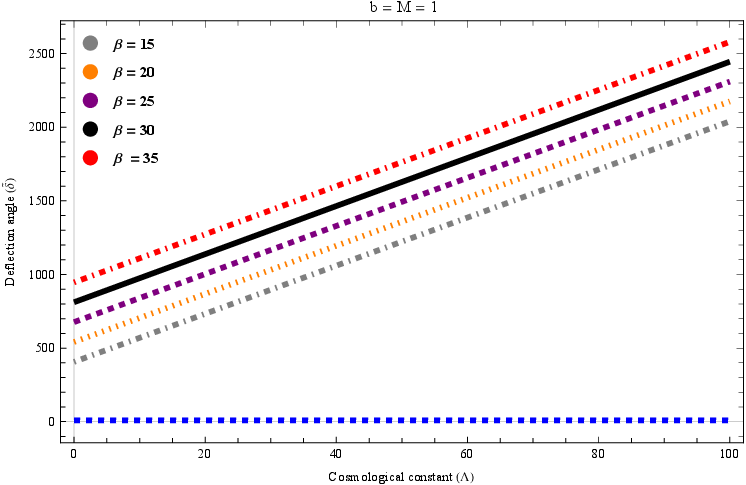}\label{3d}}
\end{array}$
\end{center}
\caption{The graph of deflection angle ($\tilde{\delta}$) with respect to cosmologival constant ($\Lambda$) by changing impact parameter $b$ with fixed $\beta = M = 1$ [(a)-(b)] and by changing $\beta$ with fixed $b = M = 1$ [(c)-(d)].}
\label{fig3}
\end{figure}

\subsection{Effect of cosmological constant (\texorpdfstring{$\Lambda$}{Lg}) on deflection angle (\texorpdfstring{$\tilde{\delta}$}{Lg})}
We present the effect of cosmological constant $\Lambda$ on the deflection angle in Fig.\ref{fig3}. Taking a look at the upper panel, we see that the deflection angle decreases (increases) with cosmological constant for small (large) variations of impact parameter $b$. An analogous result can be found in Ref. \cite{lambda}. On the other hand, the lower panel reflects that the deflection angle always increases with $\Lambda$ irrespective of $\beta$.

\subsection{Effect of real constant (\texorpdfstring{$\beta$}{Lg}) on deflection angle (\texorpdfstring{$\tilde{\delta}$}{Lg})} 
\begin{figure}[ht]
\begin{center}  
$\begin{array}{cccc}
\subfigure[]{\includegraphics[width=0.5\linewidth]{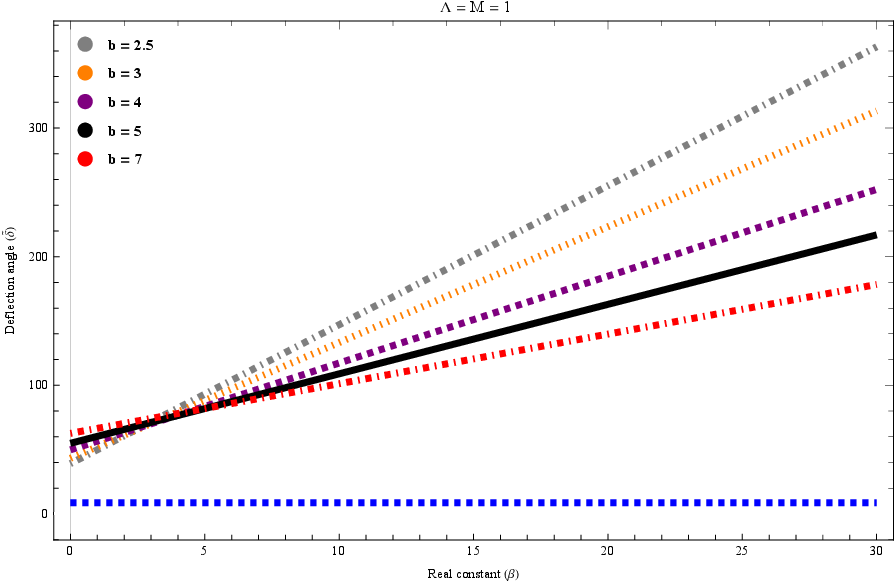}
\label{4a}}
\subfigure[]{\includegraphics[width=0.5\linewidth]{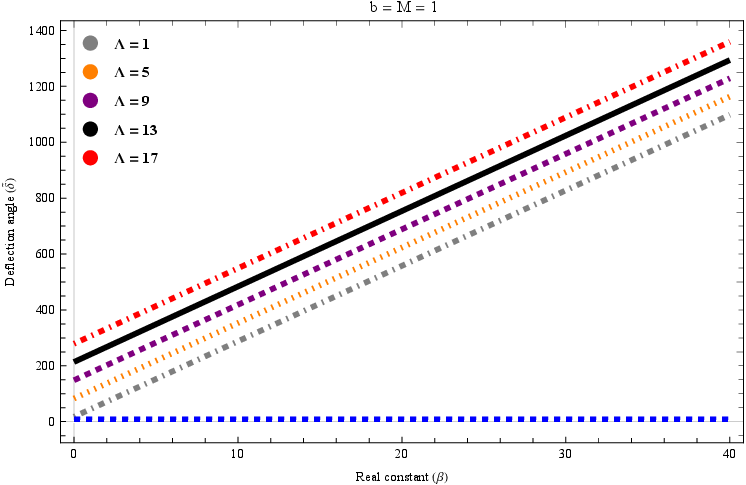}\label{4b}}  
\end{array}$
\end{center}
\caption{Plot of $\tilde{\delta}$ with respect to $\beta$ by changing impact parameter $b$ for fixed $\Lambda = M = 1$ [(a)] and by changing $\Lambda$ but $b = M = 1$ [(b)].}
\label{fig4}
\end{figure}

The impact of real constant ($\beta$) on deflection angle ($\tilde{\delta}$) is reported in Fig. \ref{fig4}. The plot tells us that $\tilde{\delta}$ is an increasing function of real constant $\beta$. Here, $\tilde{\delta}$ decreases and kepp increasing for increasing $b$ and $\Lambda$ respectively.

Meanwhile, the blue dotted curve in all the figures shows the behavior of the deflection angle for the Schwarzschild black hole case when we take the condition as $\Lambda=\beta=0$. The above graphical analysis resembles that the deflection angle is seen to be vividly influenced by parameters like $\Lambda$ and $\beta$. Due to the presence of these parameters, the deviating curve illustrates that the deflection angle for a static black hole in $f(R)$ gravity will be more than the Schwarzschild black hole case. Therefore, the presence of these extra parameters appears to be augmenting the deflection angle.
 
\section{Hawking radiation}\label{sec5}
This section aims to explore the calculation of the Hawking temperature of the static black hole in $f(R)$ gravity by employing a topological technique based on GBT. This topological technique helps us to calculate the black hole temperature by using 2-dimensional Euler characteristic $\chi$ and GBT \cite{f13,f14}. For this purpose, the line element for a 4-dimensional static spherically symmetric black hole in $f(R)$ gravity is given by \cite{f1,f15}
\begin{equation}\label{52}
ds^2=-B(r)dt^2+\frac{dr^2}{B(r)}+r^2(d\theta^2+sin^2\theta d\phi^2)
\end{equation}
By using the Wick rotation ($\tau=it$, $\theta=\pi/2$), we can get the 2-dimensional Euclidean metric from the 4-dimensional spherically symmetric metric as \cite{f16}
\begin{equation}\label{53}
ds^2=-B(r)dt^2+\frac{dr^2}{B(r)}
\end{equation} 
The event horizon radius of a static black hole in $f(R)$ gravity is estimated as
\begin{equation}\label{54}
r_{h}=\frac{\beta}{\Lambda}+\frac{2^{\frac{1}{3}}(-9\beta^2-9\Lambda)}{3\Lambda\Big(\psi+\sqrt{4(-9\beta^2-9\Lambda)^3+\psi^2}\Big)^{\frac{1}{3}}}+\frac{\Big(\psi+\sqrt{4(-9\beta^2-9\Lambda)^3+\psi^2}\Big)^{\frac{1}{3}}}{3\times2^{\frac{1}{3}}\Lambda}
\end{equation}
where $\psi = 54\beta^3+81\beta\Lambda-162M\Lambda^2$. The formula to compute the Hawking temperature of this black hole is written as follows \cite{f13,f17} 
\begin{equation}\label{55} 
T_{H}=\frac{1}{4\pi\chi}\int_{r_{h}}\sqrt{g}\mathcal{R}dr
 \end{equation}
where, $g$ denotes the determinant of Eq. \ref{54} and its value is 1, $\chi = 1$ is the Euler characteristic ans $r_{h}$ is the radius of the event horizon. The Ricci scalar $\mathcal{R}$ is obtained as \cite{hww},
\begin{equation}\label{56} 
\mathcal{R}=\frac{4M}{r^3}+\frac{2}{3}\Lambda
 \end{equation} 
After plugging the values of $g$, $R$ and then integrating Eq. \ref{54} along the event horizon, we get the Hawking temperature $T_{H}$ of the static black hole in $f(R)$ gravity as
\begin{eqnarray}\label{57a}
T_{H}=\frac{M}{2\pi r_{h}^2}-\frac{\Lambda}{6\pi}r_{h}
\end{eqnarray}
However, the standard form of the Hawking temperature of a static black hole in $f(R)$ gravity using horizon function ($T_{H}=\frac{f'(r_{h})}{4\pi}$) is
\begin{eqnarray}\label{57b}
T_{H}=\frac{M}{2\pi r_{h}^2}-\frac{\Lambda}{6\pi}r_{h}+\frac{\beta}{4\pi}
\end{eqnarray}
which is consistent with the result documented in Ref. \cite{hw}. Comparing \ref{57a} and \ref{57b}, one will find that the temperature calculated from GBT is quite different from the standard form of Hawking temperature for a static black hole in $f(R)$ gravity. Therefore, it seems that the GBT method can not determine the correct Hawking temperature for the static black hole in $f(R)$ gravity. However, it is worth mentioning that there exists an integral constant when one performs the integral \ref{55}. We can add, by keeping this in mind, an integral constant $\frac{\beta}{4\pi}$ into Eq. \ref{57a}. Consequently, we will get the
correct Hawking temperature \ref{57b} for the static black hole in $f(R)$ gravity.

Now, plugging the value of \ref{54} in \ref{57b}, we get
\begin{eqnarray}\label{57c} 
T_{H}&=&\frac{M}{2\pi}\Bigg(\frac{\beta}{\Lambda}+\frac{2^{\frac{1}{3}}(-9\beta^2-9\Lambda)}{3\Lambda\Big(\psi+\sqrt{4(-9\beta^2-9\Lambda)^3+\psi^2}\Big)^{\frac{1}{3}}}+\frac{\Big(\psi+\sqrt{4(-9\beta^2-9\Lambda)^3+\psi^2}\Big)^{\frac{1}{3}}}{3\times2^{\frac{1}{3}}\Lambda}\Bigg)^{-2}\nonumber\\
&-&\frac{\Lambda}{6\pi}\Bigg(\frac{\beta}{\Lambda}+\frac{2^{\frac{1}{3}}(-9\beta^2-9\Lambda)}{3\Lambda\Big(\psi+\sqrt{4(-9\beta^2-9\Lambda)^3+\psi^2}\Big)^{\frac{1}{3}}}+\frac{\Big(\psi+\sqrt{4(-9\beta^2-9\Lambda)^3+\psi^2}\Big)^{\frac{1}{3}}}{3\times2^{\frac{1}{3}}\Lambda}\Bigg)\nonumber\\
&+&\frac{\beta}{4\pi}
\end{eqnarray} 
where $\psi = 54\beta^3+81\beta\Lambda-162M\Lambda^2$. It is evident that the Hawking temperature calculated in Eq. \ref{57c} depends on parameters like the mass of the black hole $M$, cosmological constant $\Lambda$, and real constant $\beta$. If we take the limit $\Lambda\rightarrow 0$, $\beta\rightarrow 0$ in Eq. \ref{57c}, the calculated Hawking temperature of this black hole reduces to the Hawking temperature of the Schwarzschild black hole, $T_{H}^{Sch}=\frac{1}{8M\pi}$ (see Ref. \cite{ht}). 
\begin{figure}[ht]
\begin{center}  $\begin{array}{cccc}
\subfigure[]{\includegraphics[width=0.5\linewidth]{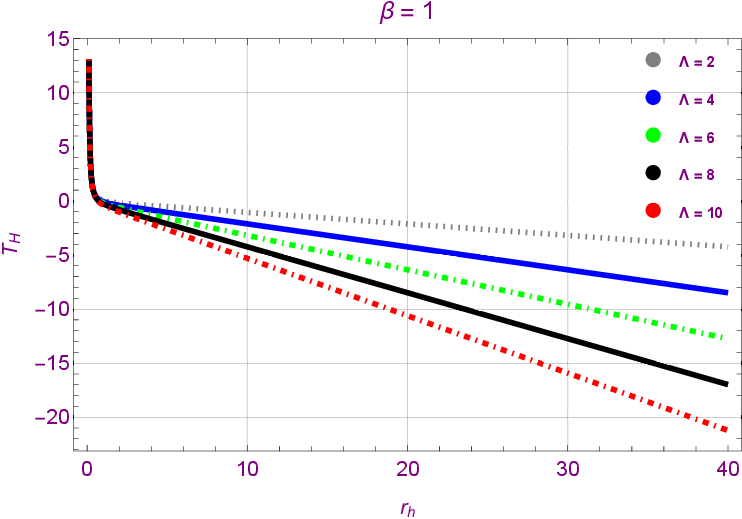}
\label{tha}}
\subfigure[]{\includegraphics[width=0.5\linewidth]{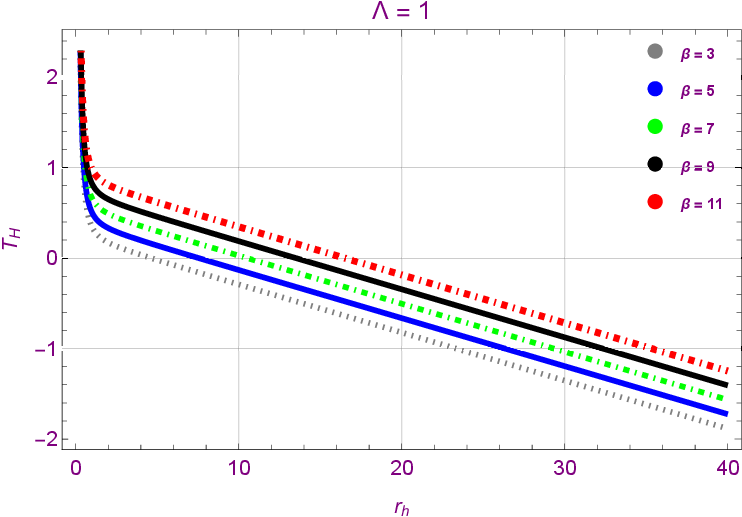}\label{thb}} \end{array}$
\end{center}
\caption{Plot of $T_{H}$ with respect to $r_{h}$ for varying $\Lambda$ (a) and $\beta$ (b). Here, $M=1$.}
\label{figth}
 \end{figure}

\begin{figure}[ht]
\begin{center}  $\begin{array}{cccc}
\subfigure[]{\includegraphics[width=0.5\linewidth]{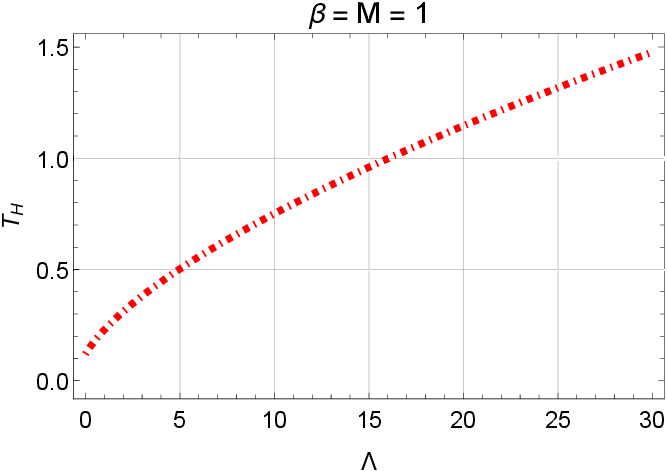}
\label{5a}}
\subfigure[]{\includegraphics[width=0.5\linewidth]{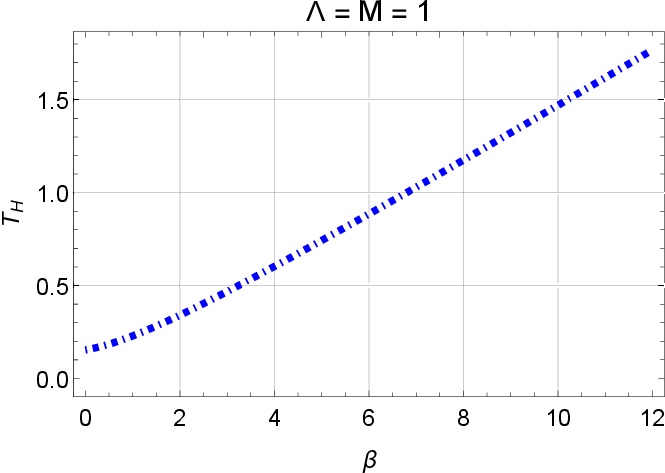}\label{5b}} \\
\subfigure[]{\includegraphics[width=0.5\linewidth]{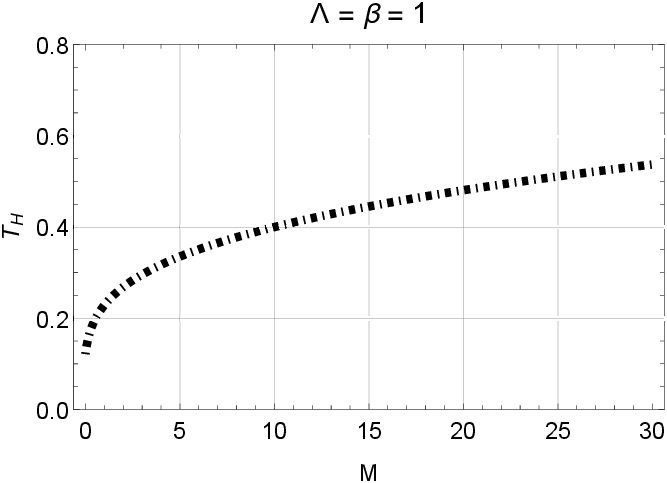}\label{5c}}  
\end{array}$
\end{center}
\caption{Plot of $T_{H}$ with respect to $\Lambda$ for fixed $\beta = M = 1$ (a), with respect to $\beta$ for fixed $\Lambda = M = 1$ (b) and with varying $M$ for fixed $\Lambda=\beta=1$ (c).  
 }
\label{fig5}
\end{figure}

To inspect the nature of Hawking temperature graphically, we present the graph of Hawking temperature $T_{H}$ with respect to $r_{h}$ (Eq. \ref{57b}) in Fig.\ref{figth} and the graph of $T_{H}$ with respect to $\Lambda$, $\beta$ and $M$ (Eq. \ref{57c}) is presented in Fig.\ref{fig5}. Fig. \ref{figth} illustrates that $T_{H}$ decreases for an increasing $r_{h}$. In addition to that, $T_{H}$ increases for decreasing $\Lambda$ and increasing $\beta$. However, we noticed from Fig.\ref{fig5} that $T_{H}$ is an increasing function of $\Lambda$, $\beta$ and $M$ and takes positive values only. This resembles that static spherically symmetric $f(R)$ black hole is always \textit{stable} when thermal fluctuation is absent. This analysis is compatible with the result elucidated in Ref. \cite{hw,qnm11}.

Moreover, further study on the first-order corrected temperature for this black hole due to small statistical fluctuations around equilibrium and also the effect of correction parameter on Hawking temperature has been elaborated in Ref. \cite{hw}.
\section{Bound on greybody factor of a static black hole in \texorpdfstring{$f(R)$}{Lg} gravity}\label{sec6}
This section is obedient to inspecting the greybody bound for static spherically symmetric $f(R)$ black hole. According to the GR, black holes can absorb all radiation and matter in the periphery of the event horizon and ejaculate Hawking radiation that depends on the mass of the black hole sustaining the quantum information of the devouring particles \cite{i45,ggg1}. The descending aspects of a radiating black hole are measured by the “greybody factor”: the terminology emphasizes
the transit from the black body behavior. Moreover, it is a function of quasinormal mode frequency ($\omega$) and angular momentum ($\it{l}$). However, the amount of deviation in the emission spectrum of a black hole from the spectrum of a perfect black body can be determined by the greybody factor. The main libellee for this inconsequence is the scattering of the radiation due to the geometry of the black
hole itself connecting to the radiated (captured) particles that decline into the extreme force. As a consequence, the spectrum emitted by the black hole will deviate and can be rendered as the
“greybody factor” \cite{ggg}. There are many strategies to find the “greybody factor” like the matching technique and WKB approximation \cite{g2}. In this present work, we will consider the technique that doesn’t put such approximations, for example, a rigorous lower bound on the “greybody factor” \cite{g3,g4,g5,g6,g7,g8,g10,g11}.

In other words, the “greybody factor” or the transmission probability in black hole physics is a quantity related to the quantum nature of a black hole which corrects the Planckian spectrum. This “greybody factor” can determine the emissivity of the given black hole solution (non-perfect blackbody). Here, we calculate the rigorous bound of the transmission probability of the static black hole in $f(R)$ gravity. Now, the general bounds of the “greybody factor” can be stated as \cite{i53,g5}
 \begin{equation}\label{58}
T \geq \operatorname{sech}^{2}\left(\frac{l}{2 \omega} \int_{-\infty}^{\infty} \mathcal{V}(r) d r_{*}\right),
\end{equation}
where, $\mathcal{V}(r)$=$\frac{\sqrt{(h^\prime)^2+(\omega^2-V-h^2)^2}}{2h}$, $r_{*}$ is the  tortoise coordinate and $\omega$ denotes the QNM frequency. Now, for $h=\omega$, Eq. \ref{58} takes the form
\begin{equation}\label{58a}
T \geq \operatorname{sech}^{2}\left(\frac{l}{2 \omega} \int_{-\infty}^{\infty} V(r) d r_{*}\right),
\end{equation}

Now, we study the Schr$\ddot{o}$dinger-like Regge-Wheeler equation for angular momentum $\it{l}$ and calculate rigorous bounds on the greybody factors. Let us introduce the Regge-Wheeler equation, which dominates the modes for a wave function $\psi(r)$ anticipating the position of an electron to its wave amplitude, as
\begin{equation}\label{59}
\left(\frac{d^{2}}{d r_{*}^{2}}+\omega^{2}-V(r)\right) \psi(r)=0,
\end{equation}
where, 
\begin{equation}
d r_{*}=\frac{1}{B(r)} d r ,\label{tor}
\end{equation}
in which $r_{*}$ is known as tortoise coordinate\footnote{The purpose of tortoise coordinate is to evolve to infinity in a way that palmist the singularity of the metric under perscrutation.} and potential $V(r)$ in $4D$ is given as
\begin{equation}\label{61}
V(r)=B(r)\Big[\frac{B^{\prime}(r)}{r}+\frac{l(l+1}{r^2}\Big].
\end{equation}
The potential for a static black hole in $f(R)$ gravity is derived as
\begin{equation}\label{61a}
V(r)=\Big(1-\frac{2M}{r}+\beta r-\frac{1}{3}\Lambda r^2\Big)\Bigg[\frac{\Big(\frac{2M}{r^2}+\beta-\frac{2}{3}\Lambda r\Big)}{r}+\frac{l(l+1}{r^2}\Bigg].
\end{equation}
This reincarnates the potential energy of the particle as long as there present an external force (or, field). The information about the scattering states, the wavefunction, the rigorous bound, etc. can be evolved from this quantity.

In order to discuss the bound value of the greybody factor, we first write the expression  of \ref{58a} with the help of \ref{tor} as
\begin{equation}\label{62}
T \geq \operatorname{sech}^{2}\left(\frac{l}{2 \omega} \int_{-\infty}^{\infty} \frac{V(r) d r}{B(r)}\right).
\end{equation}
The lower bound  value of the greybody factor (transmission probability) $T$ can be written as
\begin{equation}\label{63}
T\geq\operatorname{sech}^{2}\left[\frac{1}{2 \omega} \int_{r_{h}}^{\infty}\left( \frac{B^{\prime}(r)}{r}+\frac{l(l+1) }{r^{2}}\right)dr\right].
 \end{equation}
This expression for $T$ is the measurement of the probability of a particle having specific angular momentum ($\it{l}$) and energy runaway the black hole and being identified at infinity. Here,
$r_{h}$ is the event horizon radius of the black hole which affects the radiation spectrum emitted by the black hole. Because of the angularity of solving for the polynomial of order three, we derive $r_{h}$ numerically in Eq. \ref{54}. Moreover, the ample shape of the spectrum of the radiation emitted by a black hole is particularized by $\it{l}$.

For the metric function of our considered black hole, mentioned in Eq. \ref{34}, the above lower bound simplifies to 
\begin{eqnarray}
 T&\geq&\operatorname{sech}^{2}\left[\frac{1}{2 \omega}\left\{\frac{M}{r_{h}^2}-\beta \log(r_{h})+\frac{2}{3}\Lambda r_{h}+\frac{l(l+1)}{r_{h}}\right\} \right].
 \end{eqnarray}\label{64}
Substituting the value of event horizon $r_{h}$ from equation (\ref{54}), the bound on the greybody factor can be evaluated as 
 \begin{eqnarray}\label{65}
 T&\geq&\operatorname{sech}^{2}\Bigg[\frac{M}{2\omega}\Bigg(\frac{\beta}{\Lambda}+\frac{2^{\frac{1}{3}}(-9\beta^2-9\Lambda)}{3\Lambda\Big(\psi+\sqrt{4(-9\beta^2-9\Lambda)^3+\psi^2}\Big)^{\frac{1}{3}}}+\frac{\Big(\psi+\sqrt{4(-9\beta^2-9\Lambda)^3+\psi^2}\Big)^{\frac{1}{3}}}{3\times2^{\frac{1}{3}}\Lambda}\Bigg)^{-2}\nonumber\\&-&\frac{\beta}{2\omega} \log\Bigg(\frac{\beta}{\Lambda}+\frac{2^{\frac{1}{3}}(-9\beta^2-9\Lambda)}{3\Lambda\Big(\psi+\sqrt{4(-9\beta^2-9\Lambda)^3+\psi^2}\Big)^{\frac{1}{3}}}+\frac{\Big(\psi+\sqrt{4(-9\beta^2-9\Lambda)^3+\psi^2}\Big)^{\frac{1}{3}}}{3\times2^{\frac{1}{3}}\Lambda}\Bigg)\nonumber\\&+&\frac{\Lambda}{3\omega}\Bigg(\frac{\beta}{\Lambda}+\frac{2^{\frac{1}{3}}(-9\beta^2-9\Lambda)}{3\Lambda\Big(\psi+\sqrt{4(-9\beta^2-9\Lambda)^3+\psi^2}\Big)^{\frac{1}{3}}}+\frac{\Big(\psi+\sqrt{4(-9\beta^2-9\Lambda)^3+\psi^2}\Big)^{\frac{1}{3}}}{3\times2^{\frac{1}{3}}\Lambda}\Bigg)\nonumber\\&+&\frac{l(l+1)}{2\omega}\Bigg(\frac{\beta}{\Lambda}+\frac{2^{\frac{1}{3}}(-9\beta^2-9\Lambda)}{3\Lambda\Big(\psi+\sqrt{4(-9\beta^2-9\Lambda)^3+\psi^2}\Big)^{\frac{1}{3}}}+\frac{\Big(\psi+\sqrt{4(-9\beta^2-9\Lambda)^3+\psi^2}\Big)^{\frac{1}{3}}}{3\times2^{\frac{1}{3}}\Lambda}\Bigg)^{-1}\Bigg].
 \end{eqnarray}
where $\psi = 54\beta^3+81\beta\Lambda-162M\Lambda^2$. The lower bound derived in Eq. \ref{65} depends upon various parameter such as $M$, $\Lambda$, and $\beta$. In the limit $\Lambda=\beta=0$, one can also attain the expression of the greybody bound of the Schwarzschild black hole \cite{ggg}, as
\begin{equation}\label{sch}
T_{Sch} \geq \operatorname{sech}^{2}\Big(\frac{2l(l+1)+1}{8M\omega}\Big)
\end{equation}
\section{Graphical analysis of greybody factor with QNM frequency \texorpdfstring{$\omega$}{Lg}}\label{sec7}
 \begin{figure}[ht]
\begin{center} 
 $\begin{array}{cccc}
\subfigure[]{\includegraphics[width=0.5\linewidth]{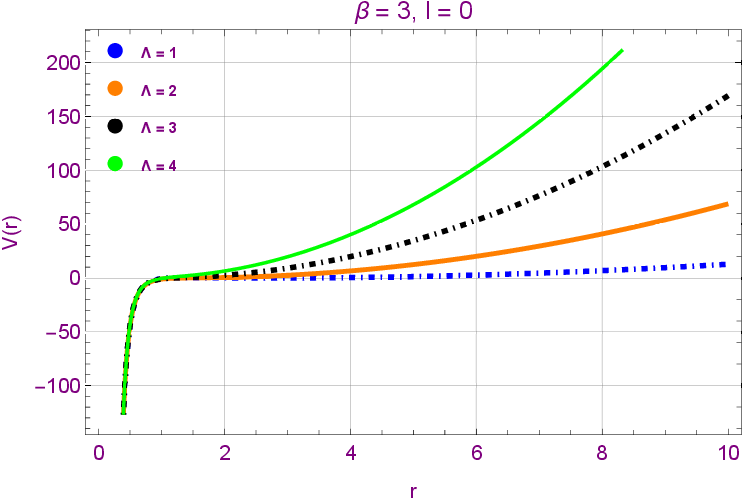}
\label{6a}}
\subfigure[]{\includegraphics[width=0.5\linewidth]{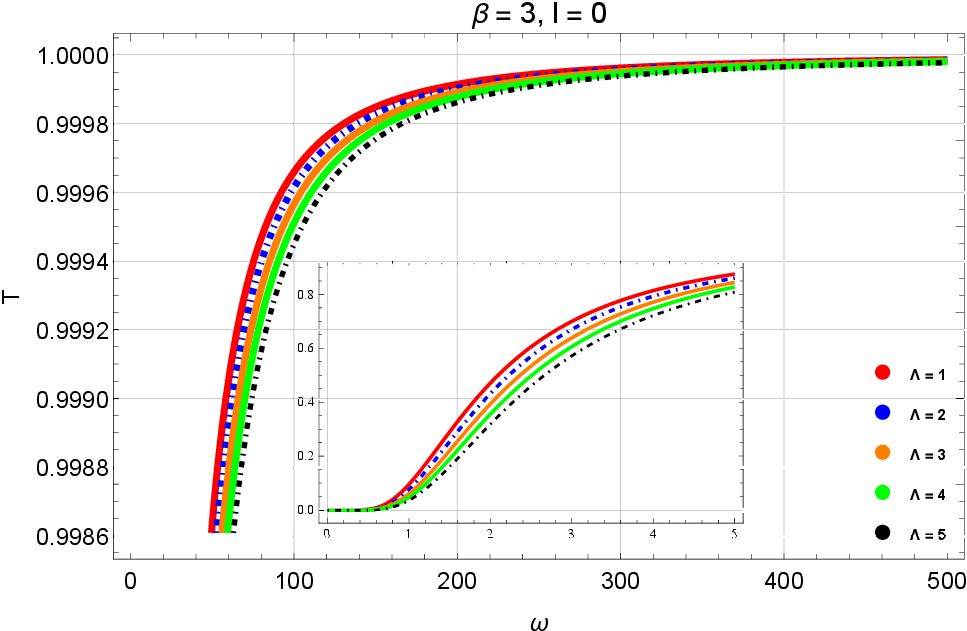}\label{6b}} \\
\subfigure[]{\includegraphics[width=0.5\linewidth]{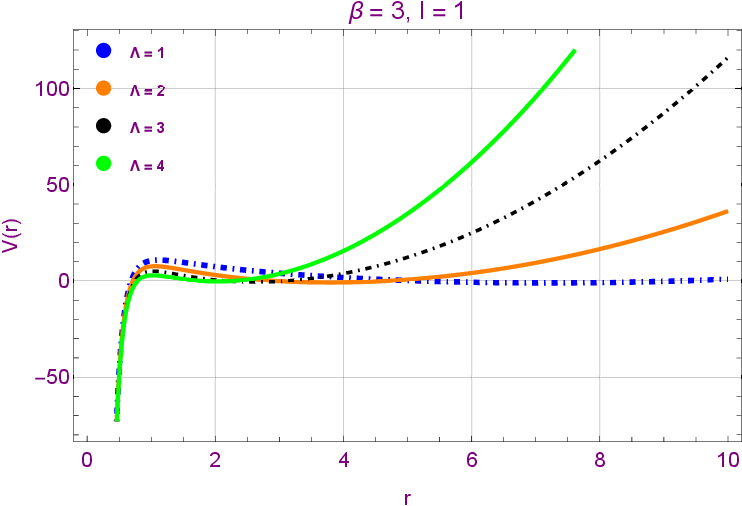}\label{6c}}
\subfigure[]{\includegraphics[width=0.5\linewidth]{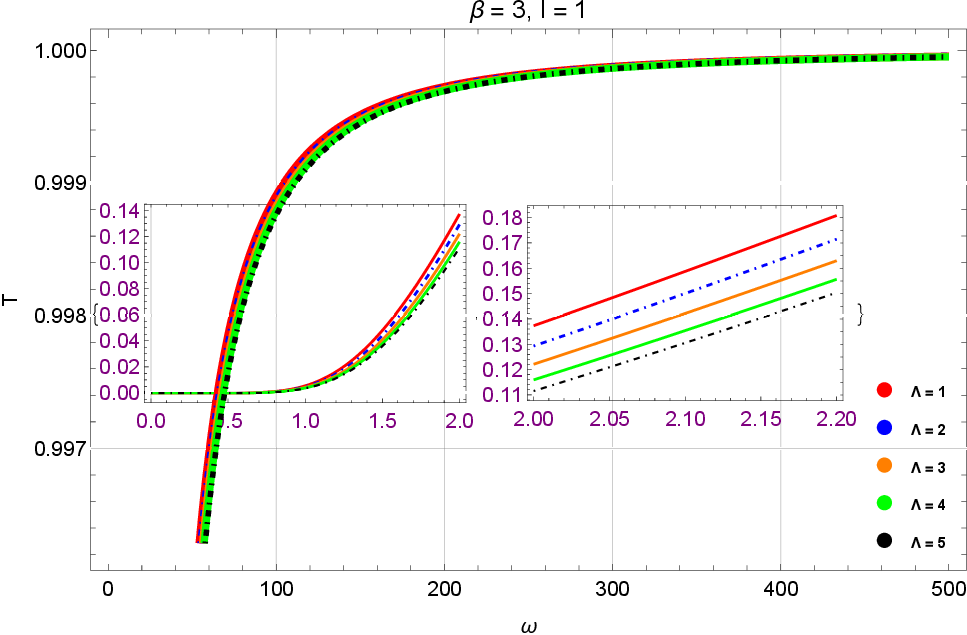} \label{6d}}
\end{array}$
\end{center}
\caption{The Left figure represents the behavior of the potential whereas the right figure corresponds to the relevant bound for varying $\Lambda$. The first panel, and second panel are for $l=0$, and $l=1$ respectively. Here, we set $M=1$ for all plots.  }
\label{fig6}
\end{figure}

\begin{figure}[ht]
\begin{center} 
 $\begin{array}{cccc}
\subfigure[]{\includegraphics[width=0.5\linewidth]{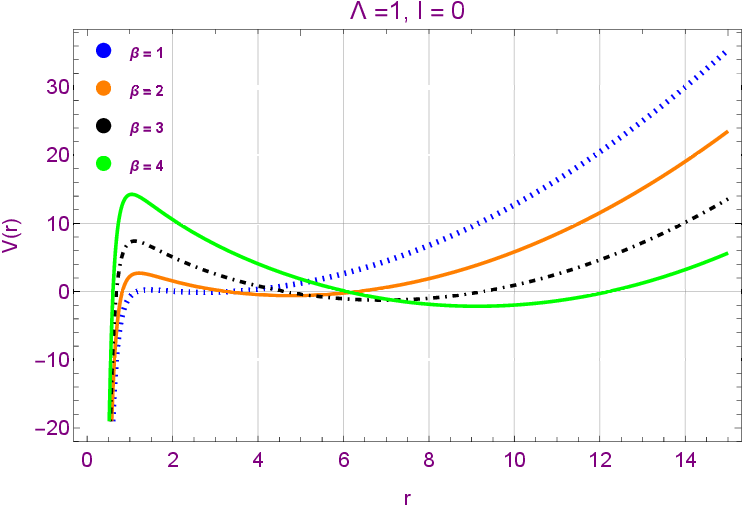}
\label{6a1}}
\subfigure[]{\includegraphics[width=0.5\linewidth]{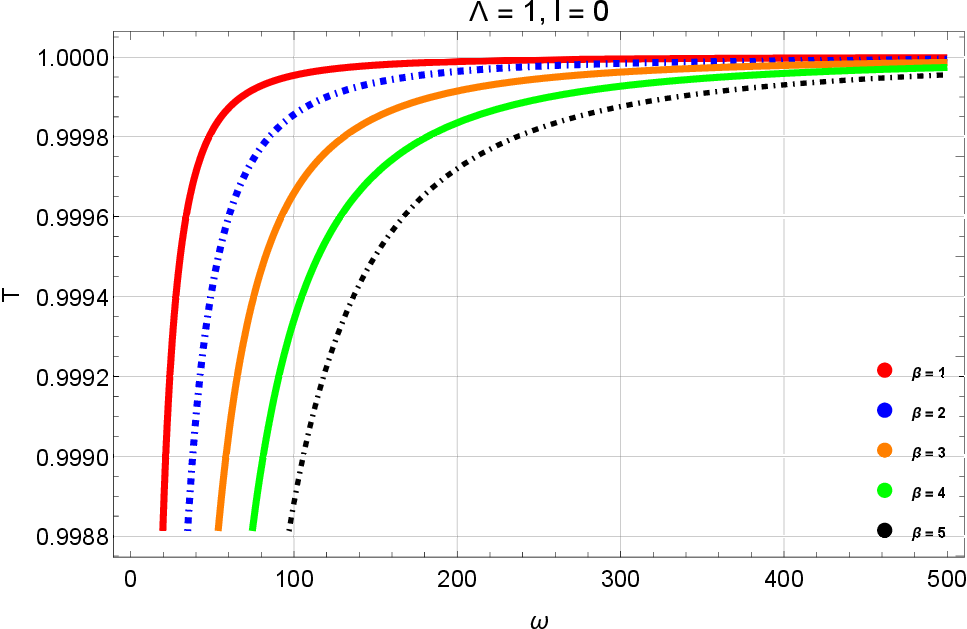}\label{6b1}} \\
\subfigure[]{\includegraphics[width=0.5\linewidth]{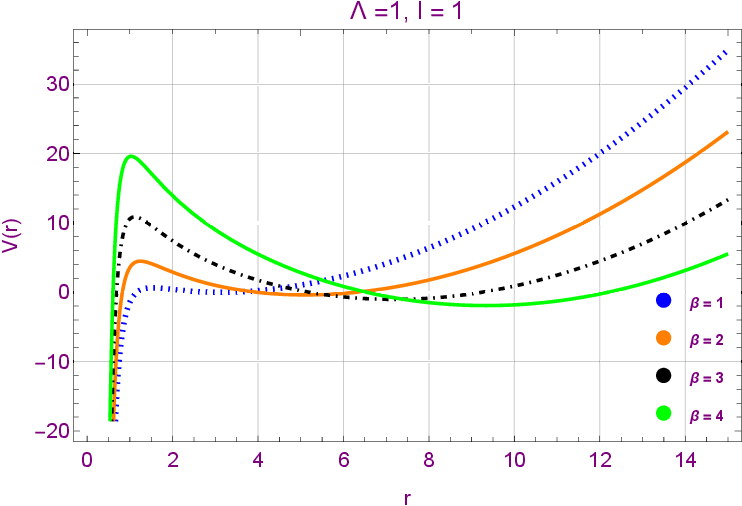}\label{6c1}}
\subfigure[]{\includegraphics[width=0.5\linewidth]{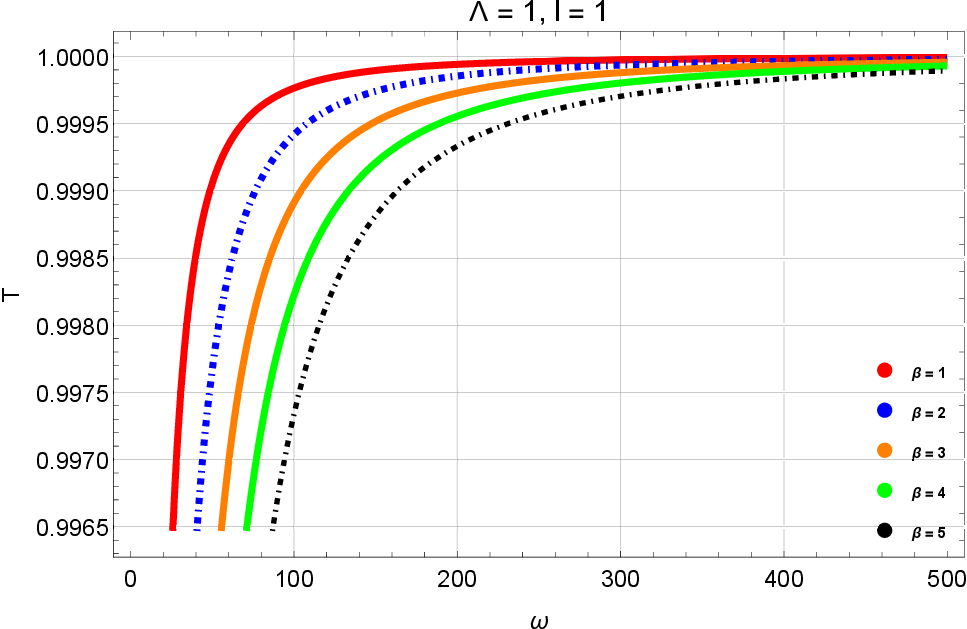} \label{6d1}}
\end{array}$
\end{center}
\caption{The Left figure represents the behavior of the potential whereas the right figure corresponds to the relevant bound for varying $\beta$. The first panel, and second panel are for $l=0$, and $l=1$ respectively. Here, we set $M=1$ for all plots.  }
\label{fig6a}
\end{figure}
In this section, we present the graphical behavior of the potential and greybody bound of a static black hole in $f(R)$ gravity. In order to do this, we consider the fixed values of mass $M$, and angular momentum $\textit{l}=0,1$ for varying cosmological constant $\Lambda$ and $\beta$. 

Figure \ref{fig6} demonstrates the graphical behavior of the greybody bound with $\omega$, and its potential with respect to $r$ with  $\Lambda$ taking values in the range $1\le \Lambda\le 5$ for fixed $\beta=3$. The domain of QNM frequency $\omega$ is taken to be $0\le \omega\le 500$. However, the inset of this figure resembles the variation of rigorous bound for lower $\omega$.

Now, taking into account Fig. \ref{6a}, we can observe the dependency of the transmission probability on the shape of potential $V(r)$ for $\textit{l}=0$. The plot shows that the potential is an increasing function of $r$ when $\Lambda$ changes. The potential increases sharply for larger $\Lambda$. We notice that the bound corresponding to $V(r)$ increases sharply and finally \textit{saturates} after a certain value of $\omega$ and it becomes 1 as long as $\omega$ approaches infinity as depicted in Fig.\ref{6b}. Nevertheless, if the value of $\Lambda$ increases then the corresponding greybody bound becomes lower which in turn makes it more difficult for the waves to be transmitted through the higher potential value as illustrated in Fig.\ref{6b}. Evidently, it is also noted that the potential and the greybody bound for $\textit{l}=1$ show similar kinds of behavior as for $\textit{l}=0$ as shown in Figs.\ref{6c} and \ref{6d}.

The graphical behavior of potential with $r$ and the corresponding greybody bound with $\omega$ for $\beta$ taking values in the range $1\le \beta\le 5$ for fixed $\Lambda=1$ is demonstrated in Fig. \ref{fig6a}. Here also the domain of $\omega$ is taken $0\le \omega\le 500$. Fig.\ref{6a1} indicates that, for $r\rightarrow 0$, the value of potential becomes high, while the potential starts decreasing from the maximum value for the large values of $r$ and further shows increasing behavior. As in Fig.\ref{fig6}, here the plot for $\it{l=0}$ delineates that the rigorous bound corresponding to potential increases sharply and displays the \textit{convergent nature} by converging to 1 after a certain value of $\omega$ as exhibited in Fig.\ref{6b1}. Furthermore, the greybody bound decreases for increasing values of $\beta$. The plot for $\it{l=1}$ parades a similar kind of demeanor as observed in Figs.\ref{6a1} and \ref{6b1}.
\section{Null geodesics of a static black hole in \texorpdfstring{$f(R)$}{Lg} gravity }\label{sec8}
This section focuses on studying the null geodesics for a static black hole in the framework of $f(R)$ gravity with due help of the Hamilton-Jacobi method. In order to achieve the shadow of this black hole, we will derive the celestial coordinates of the unstable null orbits. Nevertheless, it is worth studying null geodesics in achieving the QNMs of a black hole solution \cite{f20}. Now, we recall the following Lagrangian to explore the motion of the particle on the static black hole in $f(R)$ gravity:
\begin{equation}\label{111}
\mathcal{L} =\frac{1}{2}g_{\mu\nu} u^\mu u^\nu,
\end{equation}
in which $u^{\mu}(=\frac{dx^{\mu}}{d\lambda})$ refers to the four velocity of particle with affine parameter $\lambda$ along the geodesics.

Hence, the Lagrangian for the motion of photon of the above mentioned black hole can be written, using \ref{34}, as follows :
\begin{equation}\label{112}
2\mathcal{L}=-\Big(1-\frac{2M}{r}+\beta r-\frac{1}{3}\Lambda r^2\Big)\dot t^2+\Big(1-\frac{2M}{r}+\beta r-\frac{1}{3}\Lambda r^2\Big)^{-1}\dot r^2+r^2\dot \theta^2+r^2sin^2\theta\dot\phi^2
\end{equation}
Here the derivative with respect to the affine parameter $\lambda$ is denoted by an over dot. However, the Lagrangian has no dependency on $t$ and $\phi$ which admits two killing vectors namely $\partial_{t}$, $\partial_{\phi}$. These killing vectors give rise to the corresponding new constants of motion, namely the energy $E$ and the angular momentum $L$, given by:
\begin{equation}\label{113}
E=-p_{t}=\frac{\partial{\mathcal{L}}}{\partial{\dot t}}=-\Big(1-\frac{2M}{r}+\beta r-\frac{1}{3}\Lambda r^2\Big)\dot t
\end{equation}
and
\begin{equation}\label{114}
L=p_{\phi}=\frac{\partial{\mathcal{L}}}{\partial{\dot\phi}}=r^2sin^2\theta\dot\phi
\end{equation}
To find the restrictions of the geodesic, we can write the above constants as follows :
\begin{equation}\label{115}
\frac{dt}{d\lambda}=\dot t=\frac{E}{1-\frac{2M}{r}+\beta r-\frac{1}{3}\Lambda r^2}\;,\;\frac{d\phi}{d\lambda}=\dot\phi=\frac{L}{r^2sin^2\theta}
\end{equation}
Now, the r-part and $\theta$-part of the momentum can be defined as :
\begin{equation}\label{116}
p_{r}=\frac{\partial{\mathcal{L}}}{d\dot r}=\frac{\dot r}{1-\frac{2M}{r}+\beta r-\frac{1}{3}\Lambda r^2}\;,\;p_{\theta}=\frac{\partial{\mathcal{L}}}{\partial{\theta}}=r^2\dot\theta
\end{equation}
In order to calculate the r-part and $\theta$-part of the geodesic equations, we utilize the following relativistic Hamilton-Jacobi equation:
\begin{equation}\label{117}
\frac{dA}{d\lambda}=-\frac{1}{2}g_{\mu\nu}\frac{dA}{dx^{\mu}}\frac{dA}{dx^{\nu}}
\end{equation}
where $A$ is the Jacobi action. We can attain the Hamilton-Jacobi equation by taking into account the following ansatz \cite{f20}:
\begin{equation}\label{118}
A=\frac{1}{2}m_{\star}^2\lambda-Et+L\phi+A_{r}(r)+A_{\theta}(\theta)
\end{equation}
where $m_{\star}$ denotes the mass of the test particle, $A_{r}(r)$ and $A_{\theta}(\theta)$ are the functions of $r$ and $\theta$ respectively. For photons ($m_{\star}=0$), Eq. \ref{118} give us the following result:
\begin{equation}\label{119}
A=-Et+L\phi+A_{r}(r)+A_{\theta}(\theta)
\end{equation}

Now, by separating the values of $r$ and $\theta$,  obtain by
replacing Eq. \ref{119} into Eq. \ref{118}, we get the Carter constant ($\pm \mathcal{C}$) \cite{carter}. By plugging the values of the contravariant metric (i.e., $g^{\mu\nu}$) we obtain the following equations \cite{f21}:
\begin{equation}\label{120}
\frac{1}{\sqrt{1-\frac{2M}{r}+\beta r-\frac{1}{3}\Lambda r^2}}\frac{dr}{d\lambda}=\pm\sqrt{\mathcal{P}(r)}
\end{equation}
\begin{equation}\label{121}
r^2\frac{d\theta}{d\lambda}=\pm\sqrt{\Theta(\theta)}
\end{equation}
where 
\begin{equation}\label{121a}
\mathcal{P}(r)=\frac{E^2}{1-\frac{2M}{r}+\beta r-\frac{1}{3}\Lambda r^2}-\frac{\mathcal{C}+L^2}{r^2}
\end{equation}
\begin{equation}\label{122}
\Theta(\theta)=\mathcal{C}-\Big(\frac{L^2}{sin^2\theta}\Big)\ cos^2\theta
\end{equation}
Now, equation for $A_{r}$ can be written as follows
\begin{equation}\label{123}
\Big(\frac{dr}{d\lambda}\Big)^2+V_{eff}=0
\end{equation}
where $V_{eff}$ represents the effective potential :
\begin{equation}\label{124}
V_{eff}=-\Bigg(1-\frac{2M}{r}+\beta r-\frac{1}{3}\Lambda r^2\Bigg)\mathcal{P}(r)
\end{equation}
It is evident that the effective potential depends on the mass of the black hole $M$, real constant $\beta$, cosmological constant $\Lambda$, radius $r$, and $\mathcal{P}(r)$. Now, we introduce two new impact parameters (dimensionless), such as $\xi = \frac{L}{E}$ and $\eta=\frac{\mathcal{C}}{E^2}$. The value of $\mathcal{P}$ in terms of these two new impact parameters takes the form :
\begin{equation}\label{125}
\mathcal{P}=E^2\Bigg[\frac{1}{1-\frac{2M}{r}+\beta r-\frac{1}{3}\Lambda r^2}-\frac{\mathcal{\eta}}{r^2}\Bigg]
\end{equation}
\section{Shadow of a static black hole in \texorpdfstring{$f(R)$}{Lg} gravity}\label{sec9}
In this section, we find the shadow of a static black hole in the environments of $f(R)$ gravity. The radius of the shadow is further connected to the real part of the QNMs in the eikonal regime which will be discussed in subsection \ref{qnms}. In order to find the shadow, we will determine the unstable circular photon orbits \cite{shadow}. For this purpose, we must follow the following conditions:
$$\mathcal{P} = 0 \;and \;\mathcal{P}^{\prime} = 0$$
where prime ($\prime$) is the differentiation with respect to $r$. Substituting \ref{125} into the above conditions, we get the relation
for the photon sphere as :
\begin{equation}\label{126}
\frac{B^{\prime}(r)}{B(r)}=\frac{2}{r_{p}}
\end{equation}

Considering Eq. \ref{126} and the metric function for a static black hole in $f(R)$ gravity, $r_{p}$ is derived as 
\begin{equation}\label{1118}
r_{p}=\frac{-1+\sqrt{1+6M\beta}}{\beta}.
\end{equation}
It is observed that the above the photon sphere radius $r_{p}$ decreases when mass $M$ decreases and $\beta$ increases and it is independent of $\Lambda$. One can also check that
\begin{equation}\label{1119}
\lim_{(\beta,M)\to(0,1)}r_{p}=3 \ ,
\end{equation}
which gives the radius of an unstable photon sphere for a Schwarzschild black hole.

For the distant observer, we can measure the celestial coordinates in
the directions perpendicular ($X$) and parallel ($Y$) to the projected
rotation axis determining the apparent angular distances of the shadow image on the celestial sphere. In the present case, the celestial coordinates are given by \cite{f12a,f20,s1,s3}
\begin{equation}\label{1120}
X= \lim_{r_{0}\to\infty}\left(-r^2_{0}\sin\theta_{0}\frac{d\phi}{dr}\Big|_{r=r_{0},\theta=\theta_{0}}\right)\;,\;Y= \lim_{r_{0}\to\infty}\left( r^2_{0}\frac{d\theta}{dr}\Big|_{r=r_{0},\theta=\theta_{0}}\right).
\end{equation}
where $(r_{0}, \theta_{0})$ refers to the observer's position in the Boyer-Lindquist coordinate, the distance between an observer and black hole is denoted by $r_{0}$ and the angular coordinate (inclination angle$\footnote{Generally inclination angle is being measured between the line of sight of an observer and spinning axis of a black hole}$) of the larger observer is represented by $\theta_{0}$. In the case of null geodesic (using Eqs. \ref{115}, \ref{120}, and \ref{121}), this becomes
\begin{equation}\label{1121}
X= -\frac{\xi}{\sin\theta_{0}}\;,\;Y= \pm\sqrt{\eta-\xi^2\ cot^2\theta_{0}}\ ,
\end{equation}
which, in turn, connects the celestial coordinates with dimensionless impact parameters ($\xi,\eta$). As long as the observer lies on the equatorial plane of the mentioned black hole (i.e., $\theta_0=\frac{\pi}{2}$), the above celestial coordinates may take the following values:
\begin{equation}\label{1123}
X=-\xi,
\end{equation}
\begin{equation}\label{1124}
Y= \pm\sqrt{\eta}.
\end{equation}
Finally, the radius of the silhouette of the static black hole in $f(R)$ gravity is given by
\begin{equation}\label{1125}
R_{s}=\sqrt{X^2+Y^2} =\sqrt{\eta+\xi^2}=\frac{r_{p}}{\sqrt{B(r_{p})}}.
\end{equation}
\begin{equation}\label{1126}
R_{s}=\frac{\sqrt{3}\Big(\sqrt{1+6M\beta}-1\Big)}{\sqrt{\beta^2\Big(2\sqrt{1+6M\beta}-1\Big)-6M\Lambda\beta+2\Big(\sqrt{1+6M\beta}-1\Big)\Lambda}}.
\end{equation}

Now, the numerical values of the shadow radius $R_{s}$ for different values of a cosmological constant ($\Lambda$) and real constant $\beta$ of the static black hole in $f(R)$ gravity are presented in Table \ref{table:1} and Table \ref{table:2}, respectively.  From Table \ref{table:1} and Table \ref{table:2}, we notice that the values of the
black hole shadow radius increases with increasing $\Lambda$ however, shadow radius becomes smaller for larger real constant.
\begin{table}[h!]
\parbox{.95\linewidth}{
\centering
\begin{tabular}{ p{3cm} p{4cm} p{4cm} p{1.5cm} } 
 \hline
 $\beta$ & $r_{p}$ & $\Lambda$ & $R_{s}$ \\ [0.5ex] 
 \hline
  &  & 0.45 & 1.62617 \\ 
  &  & 0.55 & 1.70296 \\
 1 & 1.64575 & 0.65 & 1.79176 \\
  &  & 0.75 & 1.89609 \\
  &  & 0.85 & 2.02105 \\ [1ex] 
 \hline
  &  & 2 & 0.849553 \\ 
  &  & 3 & 0.974877 \\
 3 & 1.11963 & 4 & 1.17944 \\
  &  & 5 & 1.61052 \\
  &  & 5.5 & 2.13749 \\ [1ex] 
 \hline
\end{tabular}
\caption{Radius of the black hole shadow $R_{s}$ for changing $\Lambda$ with fixed $\beta$ and $r_{p}$. Here we set $M=1$.}
\label{table:1}}
\end{table}
\begin{table}[h!]
\parbox{.95\linewidth}{
\centering
\begin{tabular}{  p{3cm} p{4cm} p{4cm} p{1.5cm} } 
 \hline
 $\Lambda$ & $\beta$ & $r_{p}$ & $R_{s}$ \\ [0.65ex] 
 \hline
  & 1.5 & 1.44152 & 1.19172 \\ 
  &2  & 1.30278 & 0.966802 \\
 0.45 & 3 & 1.11963 & 0.725056 \\
  & 4 & 1.0 & 0.592349 \\
  &  5& 0.913533 & 0.506489 \\ [1ex] 
 \hline
   & 1.5 & 1.44152 & 1.25249 \\ 
  &2  & 1.30278 & 0.99841 \\
 0.65 & 3 & 1.11963 & 0.738105 \\
  & 4 & 0.592349 & 0.599401 \\
  &  5& 0.913533 & 0.510876 \\ [1ex] 
 \hline
\end{tabular}
\caption{Radius of the black hole shadow $R_{s}$ for different $\beta$ and $r_{p}$ with fixed $\Lambda$. Here, $M=1$.}
\label{table:2}}
\end{table}
\subsection{Graphical analysis of shadow}
\begin{figure}[ht]
\begin{center} 
 $\begin{array}{cccc}
\subfigure[]{\includegraphics[width=0.5\linewidth]{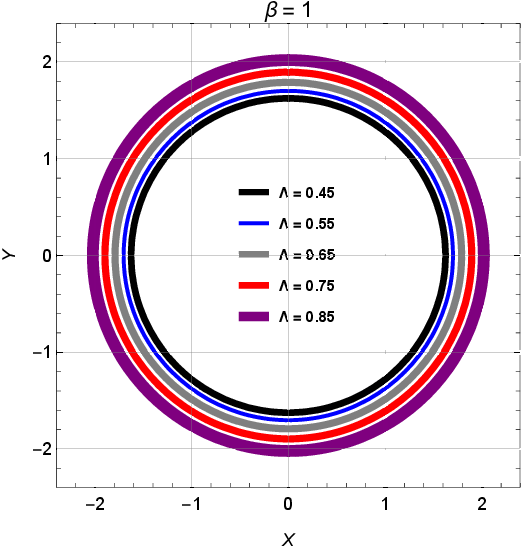}
\label{7a}}
\subfigure[]{\includegraphics[width=0.5\linewidth]{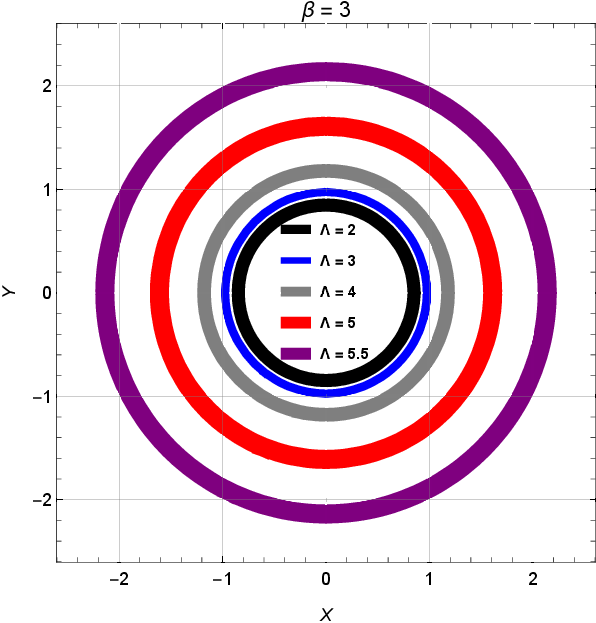}\label{7b}} 
\end{array}$
\end{center}
\caption{Black hole shadow in the celestial plane for different values of cosmological constant $\Lambda$. Here, $M=1$.  }
\label{fig7}
\end{figure}

\begin{figure}[ht]
\begin{center} 
 $\begin{array}{cccc}
\subfigure[]{\includegraphics[width=0.5\linewidth]{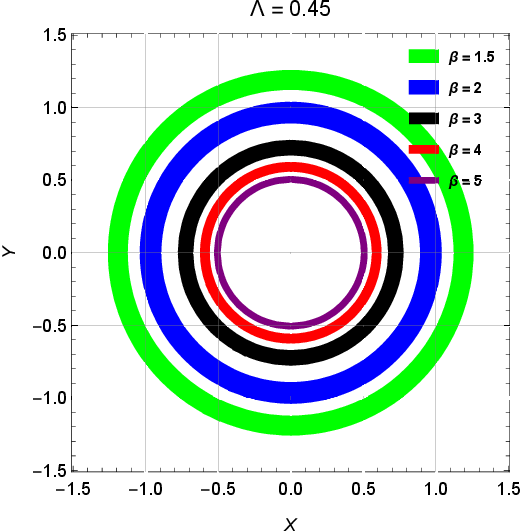}
\label{8a}}
\subfigure[]{\includegraphics[width=0.5\linewidth]{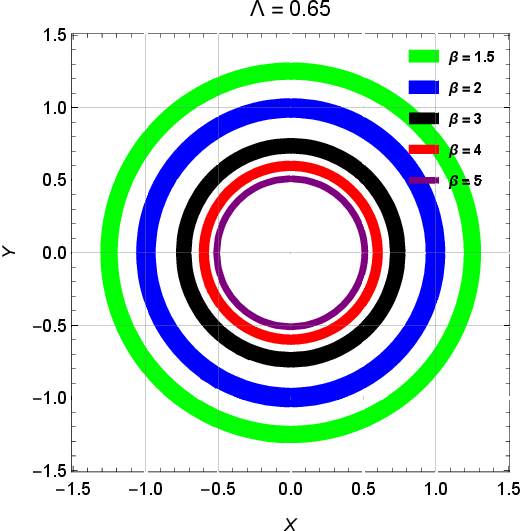}\label{8b}} 
\end{array}$
\end{center}
\caption{Black hole shadow in the celestial plane for different values of $\beta$ with $M=1$. }
\label{fig8}
\end{figure}

We present the graph of black hole shadow for different values of cosmological constant and real constant in Figs.\ref{fig7} and \ref{fig8}, respectively. In Fig.\ref{7a}, we fix the parameter $\beta=1$ whereas in Fig.\ref{7b}, we set $\beta=3$. However, in Fig.\ref{8a}, we set $\Lambda=0.45$ and in Fig.\ref{8b}, we consider  $\Lambda=0.65$. The graph displays that the shape of the shadow is perfect
circle and its shadow radius increase with an increase in cosmological constant\footnote{A similar result was found in Ref. \cite{lambda}.} while the shadow radius decreases with an increase in real constant. This explains that the gravitational field gets stronger for higher and lower values of cosmological constant and real constant respectively. The graphical analysis also justifies the numerical values shown in  Table \ref{table:1} and Table \ref{table:2}. The effect of parameters such as $\Lambda$, $\beta$, and $M$ on the black hole shadow radius is clearly presented in the next subsection.

\subsection{Effect of parameters on shadow radius in non-plasma medium}
This subsection intends to explore the intricate interplay between the shadow radius $R_{s}$ and its dependency on various parameters ($\Lambda$, $\beta$, and $M$), considering the non-plasma medium. Now, in this medium, the shadow radius for the static black hole in $f(R)$ gravity takes the following form:
\begin{equation}\label{1127}
R_{s}=\frac{r_{p}}{\sqrt{B(r_{p})}}=\frac{\sqrt{3}\Big(\sqrt{1+6M\beta}-1\Big)}{\sqrt{\beta^2\Big(2\sqrt{1+6M\beta}-1\Big)-6M\Lambda\beta+2\Big(\sqrt{1+6M\beta}-1\Big)\Lambda}}.
\end{equation} 

\begin{figure}[ht]
\begin{center} 
 $\begin{array}{cccc}
\subfigure[]{\includegraphics[width=0.5\linewidth]{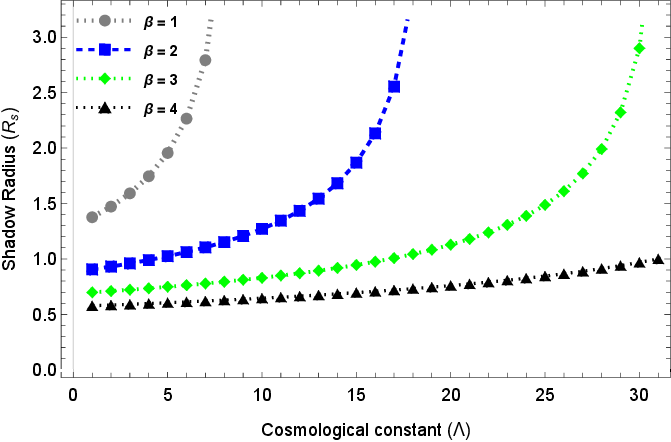}
\label{9a}}
\subfigure[]{\includegraphics[width=0.5\linewidth]{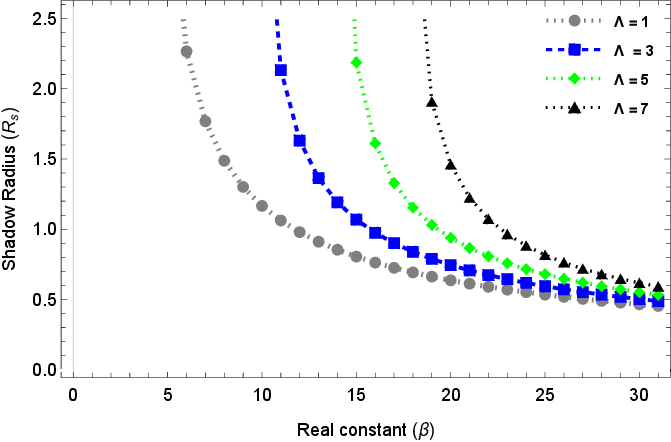}\label{9b}}\\
\subfigure[]{\includegraphics[width=0.52\linewidth]{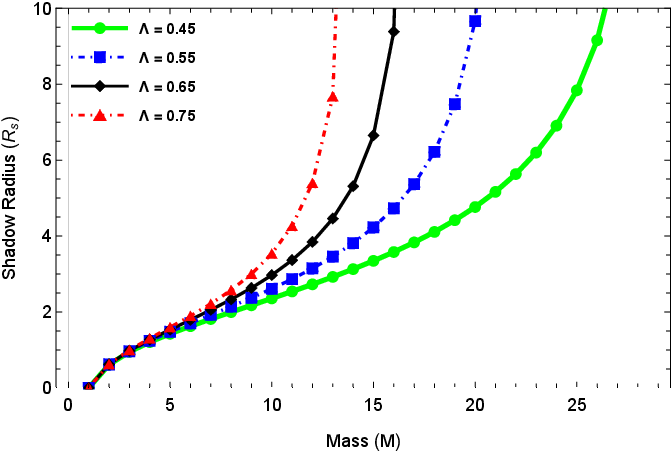}\label{9c}} 
\end{array}$
\end{center}
\caption{In \ref{9a}: variation of the shadow radius $R_{s}$ with cosmological constant $\Lambda$ for varying $\beta$ when $M=1$. In \ref{9b}: variation of the shadow radius $R_{s}$ with $\beta$ for changing cosmological constant $\Lambda$ when $M=1$. In \ref{9c}: variation of the shadow radius $R_{s}$ with $M$ for changing cosmological constant $\Lambda$ when $\beta=1$.}
\label{fig9}
\end{figure}
We see that shadow radius depends on parameters like $\Lambda$, $\beta$, and mass $M$ of the given black hole.

How $R_{s}$ depends on the parameters like $\Lambda$, $\beta$, and $M$ is depicted in Fig. \ref{fig9}. From the plot, we observe that the shadow radius is a decreasing function of real constant and an increasing function of cosmological constant and mass.
\subsection{Constraint on parameters through EHT observations}
This section aims to explore the constraints on the non-negative real constant $\beta$ associated with the static black hole in $f(R)$ gravity, assigned by the recent images of Sgr $A^{\star}$ loosened by the EHT collaboration (see Ref. \cite{EHT1}). The shadow is identified, for an observer at the distance $x$ (in Mpc) from the black hole, via its angular diameter (in µas) \cite{EHT2} as
\begin{equation}\label{EHT}
\Omega=6.191165\times 10^{-8}\frac{\xi R_{s}}{\pi x},
\end{equation}
in which $\xi$ represents the ratio of the mass of the black hole and the Sun, $x$ is the distance from the observer to the black hole while $R_{s}$ is the shadow radius mentioned in Eq. \ref{1127}. In case of the supermassive black hole Sgr $A^{\star}$, we have $\frac{M}{M_{\odot}}=\xi = 4.14\times 10^6$, $x = 8.127$ kpc \cite{EHT1,EHT3} and the observed angular diameters is $51.8 \pm 2.3$ µas \cite{EHT1} while for M $87^{\star}$, we have $\frac{M}{M_{\odot}}=\xi = 6.2\times 10^9$, $x = 16.8 $ Mpc \cite{m2} and the observed angular diameters is $42 \pm 3$ µas \cite{m2}
\begin{figure}[ht]
\begin{center}
\includegraphics[width=0.8\linewidth]{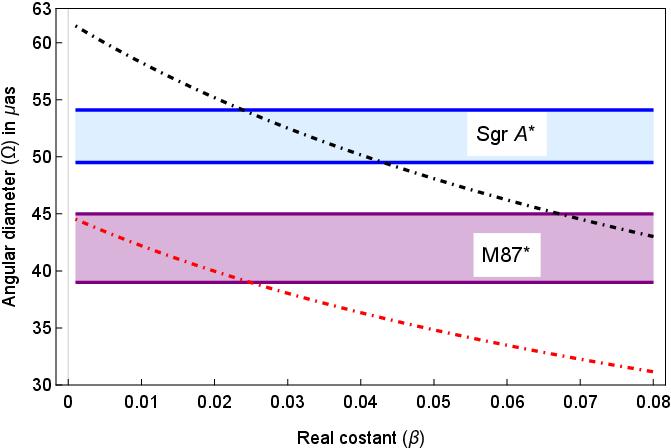}
\end{center}
\caption{The black and red curves depict the $\beta$-profiles of the angular diameter $\Omega$ for Eq. \ref{EHT}. The blue and purple regions correspond to the observed angular diameters of M$87^{\star}$ and Sgr $A^{\star}$ respectively.}
\label{fig10}
\end{figure}
In order to add some constraints on non-negative real constant $\beta$, one can utilize Eqs. \ref{1127} and \ref{EHT}. We present, with the help of  Eqs. \ref{1127} and \ref{EHT}, the bahavior of $\Omega(\beta)$ for the mentioned black hole in Fig. \ref{fig10}. The plot demonstrates that the angular diameter $\Omega$ shows a decreasing nature with $\beta$ and the $\beta$-profiles intersect with the blue and purple regions resulting in constraints on the parameter $\beta$ in the range $0 < \beta < 0.023$ for M$87^{\star}$, and $0.022 < \beta < 0.042$ for Sgr $A^{\star}$. Nevertheless, the constraint on $\beta$ giving the mean value $\beta \approx 0.011$ for M$87^{\star}$, and $\beta \approx 0.032$ for Sgr $A^{\star}$ which is consistent with the result documented in Ref. \cite{EHT5}.

\begin{figure}[ht]
\begin{center}
\includegraphics[width=0.8\linewidth]{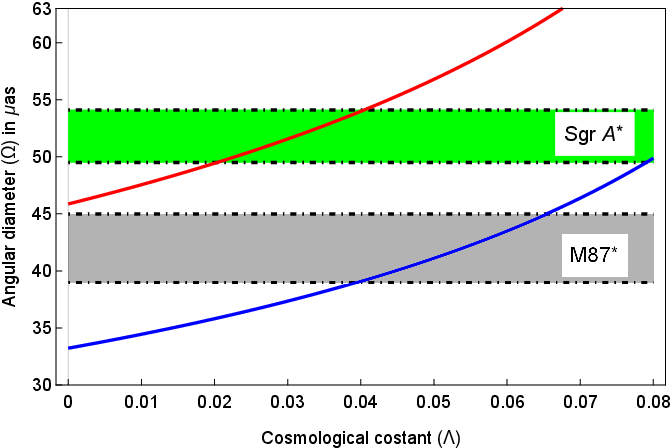}
\end{center}
\caption{The red and blue curves depict the $\Lambda$-profiles of the angular diameter $\Omega$ for Eq. \ref{EHT}. The green and brown regions correspond to the observed angular diameters of M$87^{\star}$ and Sgr $A^{\star}$ respectively.}
\label{fig11}
\end{figure}

However, Fig.\ref{fig11} helps us to constrain the value of the cosmological constant from observed results of EHT collaboration. For M$87^{\star}$, the observed domain is $0.04<\Lambda<0.065$ (with mean value $\Lambda\approx 0.053$) whereas for Sgr $A^{\star}$, the observed range is $0.02<\Lambda<0.042$ (with mean value $\Lambda\approx 0.032$).

\subsection{Correspondence between eikonal quasinormal modes and black hole shadow radius}\label{qnms}
In this subsection, we explore the correspondence between the shadow radius of the black hole and the real part of the QNMs frequency in the \textit{eikonal limit}. It has been argued by Cardoso et al. \cite{qnm2} that, in the \textit{eikonal regime}, the real part of the QNMs frequencies is connected to the angular velocity of the last unstable, circular null geodesic. However, the imaginary part of the QNMs frequencies is linked with the Lyapunov exponent that apprehends the instability time scale of the orbits. More specifically, this frequency does not depend on the initial conditions rather it depends on the details of the geometry and the type of perturbation. As per this correspondence, QNMs frequency $\omega$ can be easily calculated via the following relation
\begin{equation}\label{qnmm1}
\omega=\omega_{QNM}=\Omega_{ph}l-i\Big(n+\frac{1}{2}\Big)|\lambda^{\prime}|\ ,
\end{equation}
where $\Omega_{ph}$, and $\lambda^{\prime}$ denote the angular velocity, and the Lyapunov exponent of the unstable null geodesic respectively. Here, $n$ (= 0, 1, 2, ....) refers to the overtone number, and $\it{l}$ denotes the angular quantum number (also known as the multiple number). The angular velocity and Lyapunov exponent of the photon sphere are given, respectively, by,
\begin{equation}\label{qnmm2}
\Omega_{ph}=\frac{\sqrt{B(r_{p})}}{r_{p}}\ ,
\end{equation}
\begin{equation}\label{qnmm3}
\lambda^{\prime}=\frac{\sqrt{B(r_{p})\Big(2f(r_{p})-r_{p}^2f''(r_{p})\Big)}}{\sqrt{2}r_{p}}.
\end{equation}
Moreover, this correspondence (\ref{qnmm1}) is expected to be valid for both the static and the stationary spacetimes. On the other hand, based on Eq. \ref{qnmm1}, Stefanov et al. \cite{qnm3} pointed out a link between the QNMs and the strong deflection limit for the spherically symmetric black hole spacetime. Most recently, K. Jusufi et al. \cite{qnm4} pointed out that the real part of the QNMs frequency connected with the shadow radius via the following relation
\begin{equation}\label{qnmm4}
\omega_{R}=\lim_{l>>1}\frac{l}{R_{s}}\ ,
\end{equation}
which is accurate only in the eikonal regime having large values of angular momentum $\it{l}$ ($\it{l}>>1$). Here $R_{s}$ refers to the radius of the black hole shadow. Hence, the expression \ref{qnmm1} can take the following shape
\begin{equation}\label{qnmm5}
\omega_{QNM}=\lim_{l>>1}\frac{l}{R_{s}}-i\Big(n+\frac{1}{2}\Big)|\lambda^{\prime}|\ ,
\end{equation}

The significance of this correspondence confides on the fact that the shadow radius can be measured with due help of direct astronomical observation. Hence, instead of the angular velocity, it is more expedient to disclose the real part of the QNMs frequency in terms of the shadow radius of a black hole. Another amenity of utilizing \ref{qnmm4} is the feasibility of apprehending the shadow radius once we have determined the real part of QNMs frequency and this, in turn, does not require the help of the standard geodesic method.

\begin{figure}[ht]
\begin{center} 
 $\begin{array}{cccc}
\subfigure[]{\includegraphics[width=0.5\linewidth]{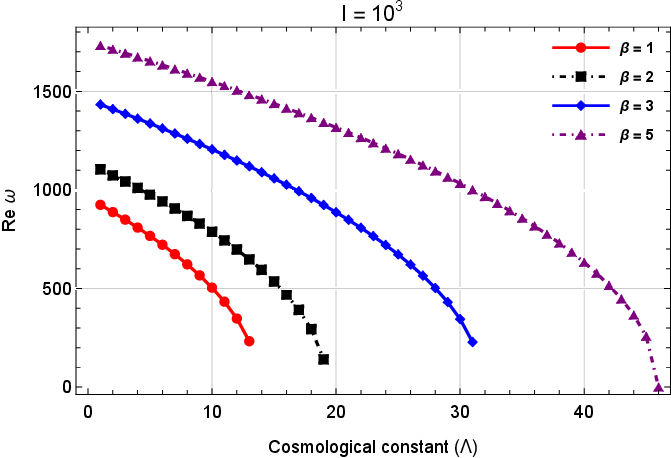}
\label{12a}}
\subfigure[]{\includegraphics[width=0.5\linewidth]{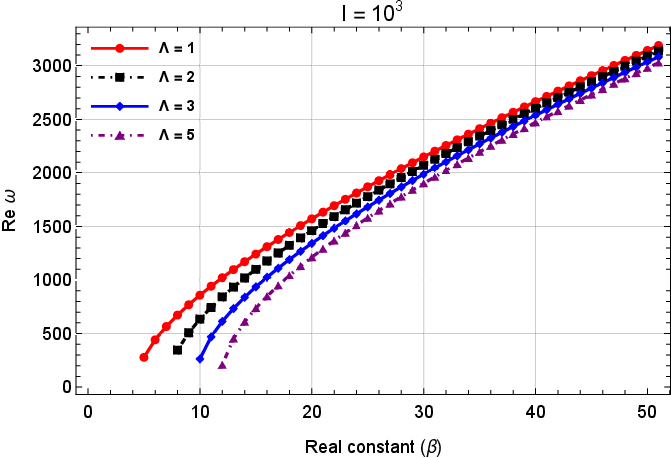}\label{12b}}\\
\subfigure[]{\includegraphics[width=0.5\linewidth]{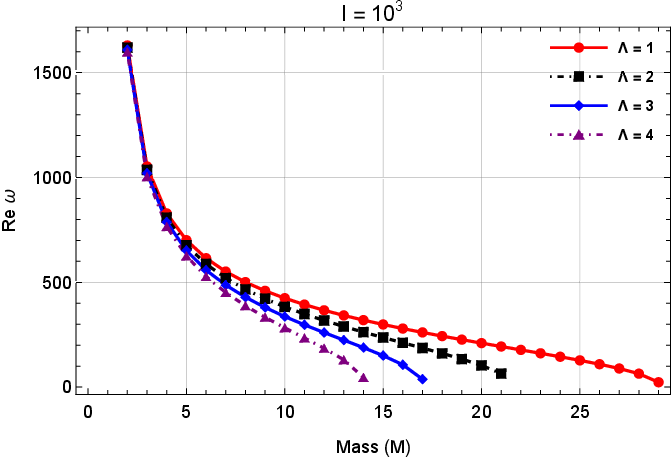}\label{12c}}
\end{array}$
\end{center}
\caption{The real part of QNMs frequency with varying $\beta$ (left figure) and varying mass (right figure) via the relation \ref{qnmm4}.}
\label{fig12}
\end{figure}

As a next step, we verify the validity of the above correspondence in our model. In order to do this, we make a graphical study (Fig.\ref{fig12}) based on Eq. \ref{qnmm4}. Evidently, we see that the real part of QNMs frequency is the decreasing function of $\Lambda$, $M$ whereas it is an increasing function of $\beta$, and this behavior of QNMs frequency is exactly opposite to the shadow radius as expected because, in the eikonal limit, the real part of QNMs frequency is inversely proportional to the shadow radius ($f_{R}\propto\frac{1}{R_{s}}$).

\begin{figure}[ht]
\begin{center} 
 $\begin{array}{cccc}
\subfigure[]{\includegraphics[width=0.5\linewidth]{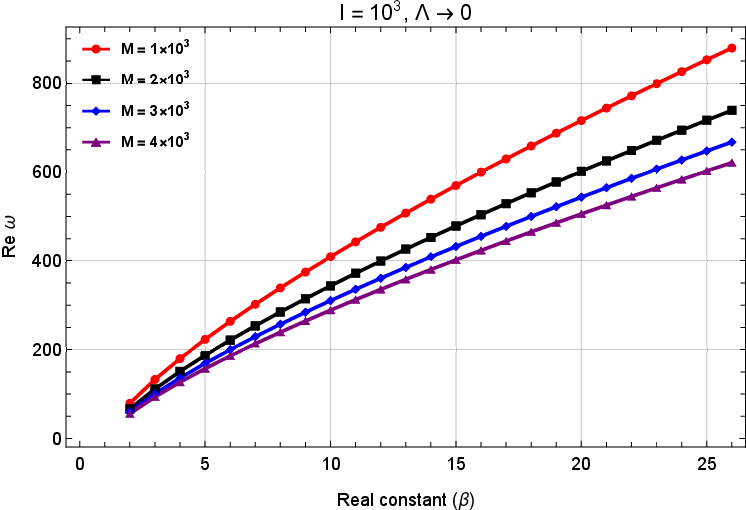}
\label{13a}}
\subfigure[]{\includegraphics[width=0.5\linewidth]{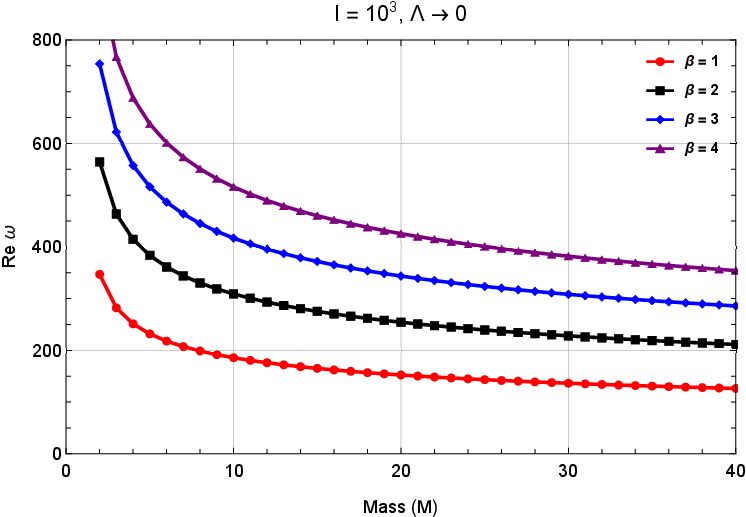}\label{13b}}
\end{array}$
\end{center}
\caption{The real part of QNMs frequency with varying $\beta$ (left figure) and varying mass (right figure) via the relation \ref{qnmm4} in cosmological constant free limit.}
\label{fig13}
\end{figure}

\begin{figure}[ht]
\begin{center} 
 $\begin{array}{cccc}
\subfigure[]{\includegraphics[width=0.5\linewidth]{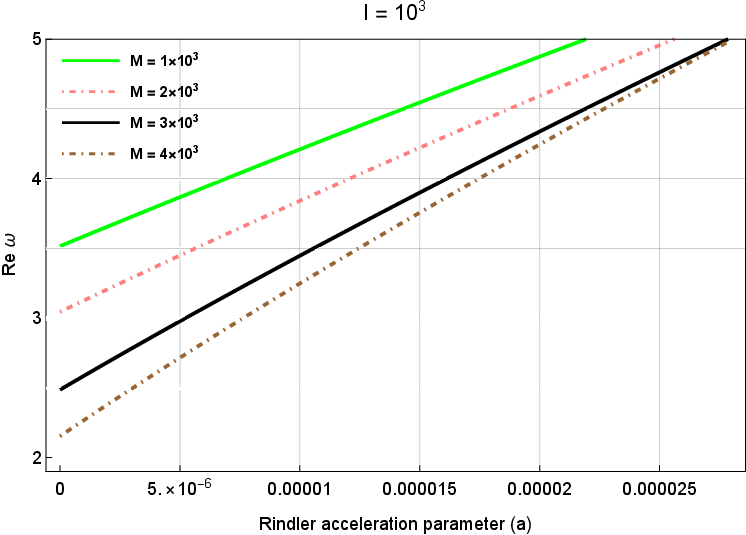}
\label{14a}}
\subfigure[]{\includegraphics[width=0.5\linewidth]{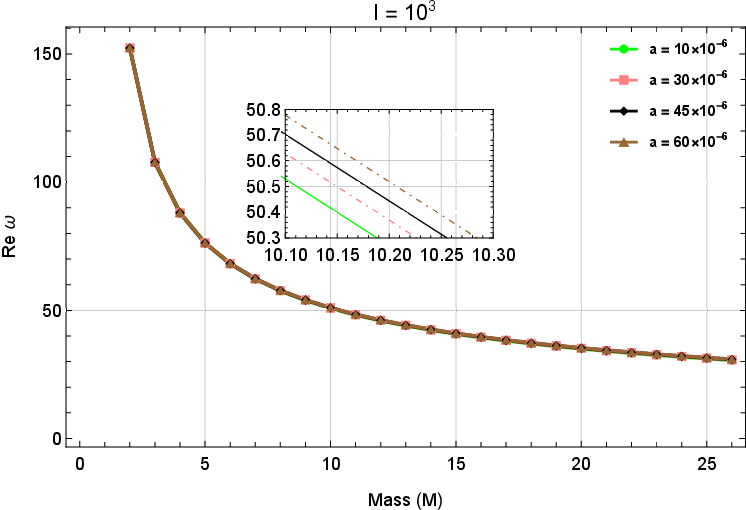}\label{14b}}
\end{array}$
\end{center}
\caption{Graphical behavior of the real part of QNMs frequency with varying $a$ (left figure) and varying mass (right figure) for GBH based on result documented in Ref. \cite{qnm10}.}
\label{fig14}
\end{figure}
It is noteworthy to mention that the metric seen in Eq. \ref{36} is the cosmological constant free version of the Grumiller black hole (GBH) or the Rindler modified Schwarzshild black hole (RMSBH) provided that real constant ($\beta$) of \ref{36} plays the same role\footnote{A recent study conducted by S. Mandal et al. \cite{qnm7} reveals that the real constant $\beta$ plays an analogous role with Rindler acceleration parameter $a$.} as that of Rindler acceleration parameter ($a$) contained in RMSBH \cite{qnm11,qnm5,qnm6,qnm7,qnm8,qnm9,qnm10,qnm12}.
Based on the study of scalar field QNMs frequency for GBH led by I. Sakalli et al. \cite{qnm10}, we draw a graph (see e.g. Fig.\ref{fig14}) of scalar field QNMs frequency with varying Rindler acceleration parameter ($a$) and mass ($M$) and notice that the real part of QNMs frequency increases (decreases) when the Rindler acceleration parameter (mass) increases. Interestingly, in the cosmological constant free cases, the nature of the real part of QNMs frequency for a static $f(R)$ black hole (see e.g. Fig.\ref{fig13}) fully corroborates with the result of GBH (see e.g. Fig.\ref{fig14} which is based on the calculation conducted Ref. \cite{qnm10}). Besides this, Ref. \cite{i32} also supports our result.
In view of this analysis, we can say that the considered model follows the correspondence between the real part of QNMs frequency and the shadow radius in the eikonal limit for scalar field perturbations.

It is expected that this correspondence in the eikonal limit could be valid for scalar, electromagnetic as well as gravitational field perturbations because they exhibit the same behavior. Nonetheless, this correspondence ceases to be valid for the gravitational field perturbations in the case of Einstein-Lovelock theory \cite{qnm13}. In GR, a similar kind of violation can be found for the charged black hole coupled with non-linear electrodynamics \cite{qnm14,qnm15}, and for the charged black hole in the framework of Eddington-inspired-Born-Infeld gravity \cite{qnm16}. Meanwhile, how this correspondence for the static $f(R)$ black hole gets affected in the presence of electromagnetic and gravitational field perturbations is not discussed here but they hold much significance. We will leave this issue for future work.

\section{Summary and final remarks}\label{sec10}
The presence of massive compact objects (such as black holes) becomes familiar through various astrophysical observations. Light ray moves through very strong gravitational fields in the vicinity of black holes. Black holes in a very special sense give amenities to investigate gravitational lensing besides the first-order weak deflection limit (which dominates most gravitational lensing). Moreover, the exploration of this new approach needs very strong technical efficiency.

In this present work, we have considered a static black hole solution in the environments of the $f(R)$ gravity model. We have calculated optical metric with due facilitation of null geodesics in the equatorial plane. In the limit of weak gravitational lensing, this optical metric introduces Gaussian optical curvature. Applying the GBT, we have estimated the deflection angle by employing the Gaussian optical curvature of the optical metric for this static $f(R)$ black hole under the consideration of a non-plasma medium. Here, we have noticed that the deflection angle of the mentioned static $f(R)$ black hole relies on several parameters like impact parameter, cosmological constant, non-negative real constant, and mass of the black hole. The presence of these excessive parameters, compared to the Schwarzschild black hole case, appears to be augmenting the deflection angle. To see the intricate interplay between these various parameters and the deflection angle, we made a graphical analysis. The graphs (Figs.\ref{fig1}-\ref{fig4}) illustrate that the lower (positive) value of the cosmological constant ($\tilde{\delta}$) enforces the deflection angle to start from a negative region and attains its maximum values by showing a peak in the positive region for lower $b$ and then increases softly with $b$ in positive region after performing declination demeanor. In contrast, for a higher (positive) valued cosmological constant, the deflection angle begins from the negative region by illustrating the increasing nature for smaller $b$ while for higher $b$, it goes to the positive region by retaining increasing tendency. The deflection
angle shows an increasing tendency with impact parameter $b$ for small values of $\beta$ and
remains positively valued in the higher region of $b$. Here, $\tilde{\delta}$ always increases for increasing $\beta$. However, for large $\beta$,
$\tilde{\delta}$ presents an asymptotical nature for small $b$, after that it increases softly with $b$ in the positive region after showing declination
behavior. For both small and large $M$, $\tilde{\delta}$ increases abruptly for
small $b$ and it finally deliberates increasing demeanor for large $b$ after performing certain
decline nature. Furthermore, the mass has an increasing (decreasing) contribution on $\tilde{\delta}$ for small (large) value of cosmological constant and large (small) value of real constant and remains positively (negatively) valued. Moreover, we noticed that the deflection angle decreases (increases) with cosmological
constant for small (large) variations of impact parameter $b$ and takes negative (positive) values only. An analogous result has been found in Ref. \cite{lambda}. Finally, $\tilde{\delta}$ shows an increasing tendency with non-negative real constant $\beta$.

In the context of GBT, we have also calculated Hawking temperature for this static $f(R)$ black hole. The resulting Hawking temperature obeys the standard form of the Hawking Temperature ($T_{H}=\frac{f^{\prime}(r_{h})}{4\pi}$). The Hawking temperature calculated in Eq. \ref{57c} depends on parameters like the mass of the black hole $M$, cosmological constant $\Lambda$, and non-negative real constant $\beta$. Moreover, If we consider the limit $\Lambda\rightarrow 0$, $\beta\rightarrow 0$ in Eq. \ref{57c}, the calculated Hawking temperature of this black hole reduces to the Hawking temperature of the Schwarzschild black hole, $T_{H}^{Sch}=\frac{1}{8M\pi}$ (see Ref. e.g. \cite{ht}). To investigate the behavior of Hawking temperature graphically, we present the graph of Hawking temperature $T_{H}$ with respect to $r_{h}$ in Figure \ref{figth} and the graph of $T_{H}$ with respect to $\Lambda$ and $\beta$ is presented in Figure \ref{fig5}. Fig. \ref{figth} illustrates that $T_{H}$ decreases for an increasing $r_{h}$. In addition to that, $T_{H}$ increases for decreasing $\Lambda$ and increasing $\beta$. However, Fig. \ref{fig5} indicates that $T_{H}$ increase with $\Lambda$, $\beta$ and $M$ and decrease with $\beta$ for fixed $\Lambda$ and takes always positive values. Hence, this concludes that a static spherically symmetric $f(R)$ black hole is always \textit{stable} in the absence of thermal fluctuations.

Furthermore, we have derived the rigorous analytic bounds of the greybody factor of the static black hole in the $f(R)$ theory of gravity and noticed that the calculated bound in Eq. (\ref{65}) depends upon various parameters such as $M$, $\Lambda$, and $\beta$. In the limit $\Lambda=\beta=0$, one can also retain the expression of the greybody bound of the Schwarzschild black hole. From the potential and bound on greybody factor graphs, the bound corresponding to potential increases sharply and then \textit{saturates} after a certain value of QNM frequency $\omega$ resulting in 1 as long as $\omega$ approaches infinity. The greybody bound decreases for both increasing $\Lambda$ and $\beta$. It is worth noting that for large
values of $\Lambda$ and $\beta$, the potential becomes higher which in turn makes it difficult for the waves to be transmitted through that potential.

Finally, we studied the shadow cast by a static $f(R)$ black hole for a distant observer by investigating the null geodesic of this black hole. Using the celestial coordinates ($X, Y$), we also derived the radius of the shadow and presented its image in the outlaying observer’s sky. We see that the shape of the silhouette is a perfect circle. We also studied the effect of parameters such as $\Lambda$, $M$, and $\beta$ on the ceremonial of the shadow radius in view of a non-plasma medium.  Our study reveals that shadow radius is a \textit{decreasing function} of \textit{real constant} $\beta$ and an \textit{increasing function} of \textit{cosmological constant} and \textit{mass} and our analysis is \textit{compatible} with the result provided in Ref. \cite{lambda}. In this regard, to constrain the dark-matter-related non-negative parameter $\beta$ and cosmological constant $\Lambda$, we estimate the theoretical angular size of the black hole shadow and perform comparisons with the observed results of EHT for M$87^{\star}$ and Sgr $A^{\star}$. We noticed that this $\beta$ lies in the range $0 < \beta < 0.023$ for M$87^{\star}$  and $0.022 < \beta < 0.042$ for Sgr $A^{\star}$ and our result is \textit{consistent} with the result provided in Ref. \cite{EHT5}. In case of $\Lambda$, for M$87^{\star}$, the observed domain is $0.04<\Lambda<0.065$ whereas for Sgr $A^{\star}$, the observed range is $0.02<\Lambda<0.042$. Moreover, the connection between the shadow radius and \textit{eikonal} QNMs frequency has been employed. In this context, we conclude that the considered $f(R)$ model follows the correspondence between the real part of QNMs frequency and the shadow radius in the \textit{eikonal limit} for \textit{scalar field perturbations}.

In future research endeavors, directing our conjecture toward expected exploration, it becomes tempting to intend the extension of these findings to encircle the effect of both the plasma and dark matter medium on the deflection angle and shadow. Moreover, in the framework of $f(R)$ gravity theory, comprehensive scrutiny such as the consequences of spin parameters on the deflection angle, shadow, and greybody bound of the black hole apprehends the pledge for manifesting us with profound insights.  Apart from this, how the correspondence between the real part of \textit{eikonal} QNMs frequency and shadow radius for the static $f(R)$ black hole gets affected in presence of \textit{electromagnetic} and \textit{gravitational field} perturbations holds much significance. We will leave this issue for future research work.

\section*{Acknowledgements}
The author is thankful to Surajit Das (Department of Physics, Cooch Behar Panchanan Barma University, Coochbehar, West Bengal, India) for various suggestions which developed the presentation of the paper. The author thanks the Editor and anonymous Referee for their constructive comments and valuable suggestions.

\section*{Declaration of competing interest}
The authors declare that they have no known competing financial interests or personal
relationships that could have appeared to influence the work reported in this manuscript.
 
\section*{Data Availability Statement}
Data sharing is not applicable to this article as no data sets were generated or analyzed during the current study.

\end{document}